\documentclass{emulateapj}
\usepackage{amssymb}
\usepackage{amsmath,bm}
\usepackage[normalem]{ulem}
\usepackage[dvips]{color}
\usepackage{subfigure}
\usepackage[english]{babel}
\usepackage{multirow}
\usepackage[mathcal]{eucal}
\usepackage{rotating}
\usepackage{graphicx}
\usepackage{dcolumn}
\usepackage{bm}
\usepackage{graphicx}
\usepackage{CJK}
\usepackage{array}

\renewcommand\sout{\bgroup \color{red} \ULdepth=-.5ex \ULset}

\begin{document}

\title{Old neutron stars as probes of isospin-violating dark matter}
\author{Hao Zheng$^1$, Kai-Jia Sun$^1$, and Lie-Wen Chen$^{*1,2}$}
\affil{$^1$ Department of Physics and Astronomy and Shanghai Key Laboratory for
Particle Physics and Cosmology, Shanghai Jiao Tong University, Shanghai 200240, China}
\affil{$^2$ Center of Theoretical Nuclear Physics, National Laboratory of Heavy Ion
Accelerator, Lanzhou 730000, China}
\altaffiltext{*}{Corresponding author (email: lwchen$@$sjtu.edu.cn)}
%
%

\begin{abstract}
Isospin-violating dark matter (IVDM), which couples differently with protons and neutrons,
provides a promising mechanism to ameliorate the tension among recent direct detection
experiments. Assuming DM is non-interacting bosonic asymmetric IVDM,
we investigate how the existence of old neutron stars limits
the DM-proton scattering cross-section $\sigma_{\rm p}$, especially the effects of
the isospin violating DM-nucleon interactions and the symmetry energy in the equation of state of
isospin asymmetric nuclear matter.
Our calculations are completely based on general relativity and especially the
structure of neutron stars is obtained by solving the Tolman-Oppenheimer-Volkoff equations
with nuclear matter equation of state constrained by terrestrial experiments.
We find that, by considering the more realistic neutron star model
rather than a simple uniform neutron sphere as usual, the $\sigma_{\rm p}$ bounds from old neutron stars
can be varied by more than an order of magnitude depending on the specific values of the
DM neutron-to-proton coupling ratio $f_{\rm n}/f_{\rm p}$, and they can be further varied by more
than a factor of two depending on the density dependence of the symmetry energy.
In particular, we demonstrate that the observed nearby isolated old neutron star PSR B1257+12
can set a very strong limit on $\sigma_{\rm p}$ for low-mass DM particles ($ \le 20 \, {\rm GeV}$) that
reaches a sensitivity beyond the current best limits from direct detection experiments and disfavors
the DM interpretation of previously-reported positive experimental results, including the IVDM.
\end{abstract}



\section{Introduction}
\label{sec:intro}

The quest for dark matter (DM) is one of the most intriguing
aspects of current research frontiers of particle physics,
astrophysics and cosmology. Cosmological observations
have provided compelling evidence for the existence
of DM. The most recent cosmological results based on {\it Planck} measurements
of the cosmic microwave background (CMB) temperature and lensing-potential power
spectra indicate that DM comprises about $27\%$ of the energy density of the
Universe which also contains about $5\%$ baryon matter and about $68\%$ dark
energy~\citep{Plk13}. On the other hand, the DM nature
and its interactions with the Standard Model particles remain unknown.

Many theories beyond the Standard Model of particle physics predict natural
candidates for DM. Among them, motivated by the fact that the DM and baryon
densities in the Universe are of the same order of magnitude, asymmetric
dark matter (ADM)~\citep{Zur13,Vol13}
has been proposed as one of the most promising classes of candidates for DM.
Having inherited a similar matter-antimatter asymmetry as baryons, the models of ADM
mainly focus on the DM mass region around a few GeV.
In this mass region, these models are favored by an excess of events over
the expected background observed in some underground DM direct
detection experiments such as CoGeNT~\citep{CoGeNT11},
DAMA~\citep{DAMA09} and CRESSTII~\citep{CRESST12} as well as the recent results
presented by the CDMS-II(Si) collaboration~\citep{CDMS13v1,CDMS13v2}.
However, these observational evidences are in strong tension with the
constraints set by some other experimental groups like
XENON100~\citep{XENON100v11,XENON100v12}, LUX~\citep{LUX13} and
SuperCDMS(Ge)~\citep{SCDMS14}.

Isospin-Violating Dark Matter (IVDM) provides a possible mechanism to
reconcile the tension among different experiments~\citep{Kur04,Giu05,Cha10,Feng11,Feng13a,Feng13b,Nag13,Vin13,Zheng14}.
Within the IVDM framework, DM is assumed to couple differently with protons
and neutrons, and this assumption of the isospin violation has been supported
by a number of theoretical works~\citep{Fra11,Cli11,Del12,He12,Gao13,Oka13} based on the
particle physics point of view. Especially, IVDM can also be asymmetric in
the model proposed by Okada and Seto in Ref.~\citep{Oka13}.

Besides the underground DM direct detection experiments in terrestrial
laboratory, the existence of old
neutron stars in our Galaxy can also provide constraints on
the scattering cross-sections between DM and nucleons.
This subject was firstly studied by Goldman and Nussinov~\citep{Gold89},
and then a lot of works have been done during the last few
years~\citep{Ber08,Lava10,Chris11,Chris12,Chris13,Yu12,Kumar13,Kumar14,Jami13,Bell13,Guv14}.
Especially for ADM, due to the asymmetry between particles
and anti-particles in the DM sector, DM annihilation is
not significant in this case. Therefore, DM can continually accumulate in the neutron star without depleting.
After being thermalized through continually scattering with the neutron star matter,
these DM particles will gather within a small thermal radius at the central region of the neutron star,
then become self-gravitating and eventually collapse into black holes
which could destroy the host neutron star.

For non-interacting fermionic DM, due to the Fermi degeneracy pressure, the
required number for the self-gravitating DM to collapse into black holes (i.e., the Chandrasekhar limit) is about
$\left(M_{\rm pl}/m_{\chi}\right)^3$~\citep{Shapiro86},
where $M_{\rm pl}$ is the Planck mass and $m_{\chi}$ is the DM mass.
Except for DM with a huge $m_{\chi}$, this number is far beyond the amount of DM particles that could be captured
by a neutron star within a few billion years inside a DM halo with density
similar to that around the earth.
On the other hand, for non-interacting bosonic DM the degeneracy pressure does not exist,
and the relevant Chandrasekhar limit for collapsing becomes much smaller, i.e.,
about $\left(M_{\rm pl}/m_{\chi}\right)^2$~\citep{Shapiro86}.
This DM amount can be easily accumulated within the living age of a typical old
neutron star if the DM-nucleon scattering cross-section is not too small.
Furthermore, for the bosonic DM, the possible existence of the Bose-Einstein condensate (BEC)
inside the neutron star could further facilitate the onset of self-gravitation,
leading to an enhancement effect on the gravitational collapse.
Once the mini black hole is formed inside the neutron star, it could subsequently
lead to the ultimate destruction of the neutron star if it consumes the neutron star matter
faster than the Hawking evaporation.
However, old nearby neutron stars have been well observed, and this thus
could result in a very severe constraint on
the scattering cross-sections of the non-interacting bosonic ADM with nucleons.
The above arguments are based on the assumption that the DM particles have no self-interactions
(except for gravity). It should be mentioned that
for fermionic DM, the required DM particle number for gravitational
collapse can be strongly reduced if an attractive
Yukawa force can appear among DM particles~\citep{Chris12,Kumar14}.
In addition, the repulsive self-interactions as well as the effect
of DM-nucleon coannihilations can also
prevent the black hole formation for bosonic DM~\citep{Kumar13,Bell13}.

In previous studies, the neutron stars have been generally assumed to be
consisted of pure neutrons with constant density throughout a sphere.
However, for a realistic neutron star, its composition is much more
complicated than a simple uniform neutron sphere~\citep{Latt04}.
That is, not only neutrons but also protons, leptons and even
some non-nucleonic degrees of freedom may appear inside the neutron star.
Furthermore, the density profiles for various constituents in the neutron star
are far beyond uniformly distributed. These variations could
become even more important and interesting when we deal with IVDM which
interacts with various components differently. For example,
in the extreme case that DM only couples to protons (as for anapole interactions~\citep{Zur10,Ho10}),
constraints on DM-nucleon cross sections set by neutron stars
would become much weaker (but non-zero) due to the fact that
the proton fraction is usually much smaller
than the neutron fraction inside a typical neutron star.

The motivation of the present work is twofold. Firstly, we
investigate how the isospin violating DM-nucleon interactions and the
symmetry energy in the equation of state (EOS) of isospin asymmetric
nuclear matter affect the extraction of the bound on the DM-proton
scattering cross-section $\sigma_{\rm p}$ from the existence of old neutron stars, by
considering, for the first time, a more realistic neutron star model
in which the neutron star is assumed to be static and composed of $\beta$-stable
and electrically neutral $n p e \mu$ matter (i.e., the so-called
conventional neutron stars) and its structure is obtained by solving
the Tolman-Oppenheimer-Volkoff (TOV) equations. Secondly, we extract
the bound on the DM-proton scattering cross-section $\sigma_{\rm p}$ from the existence
of a realistic isolated pulsar, i.e., the observed nearby old neutron
star PSR B1257+12, and compare the results with those from direct
detection experiments, especially for IVDM.
We restrict our attention on the non-interacting bosonic asymmetric IVDM within
the GeV mass region which is favored by recent DM direct detection
experiments. Our results indicate that both the isospin violating DM-nucleon interactions
and the symmetry energy can significantly affect the extraction of the bound on the
DM proton scattering cross-section $\sigma_{\rm p}$ from the existence of old neutron stars.
In particular, the observed nearby isolated old neutron star PSR B1257+12 can set a
stringent limit for low-mass DM particles that reaches a sensitivity beyond the current
best limits from direct detection experiments and excludes the DM interpretation of
all previously-reported positive experimental results, including the IVDM.

This article is organized as follows. In Sec.~{\ref{sec:model}} we describe
the main models and methods used in the present work, including a brief introduction to
the symmetry energy and neutron star structure, how to calculate the capture rates of the DM particles onto
a structured neutron star and how to determine the conditions
for the black hole formation inside neutron stars. In Sec.~{\ref{sec:results}}
we give the main results and discussions, including the
symmetry energy effects on the global properties of a typical neutron
star, the constraints on the DM-proton
scattering cross sections for various isospin-violating cases as well as
different symmetry energy cases, and the constraints on the DM-proton scattering cross sections from the
observed nearby isolated old neutron star PSR B1257+12.
We present our conclusions in Sec.~{\ref{sec:conclusion}}.
An appendix is given to describe the general relativity corrections
of DM thermalization and BEC formation in neutron stars.

\section{Models and Methods}
\label{sec:model}

\subsection{The symmetry energy and neutron star structure}
\label{subsec:EsymNStar}

The symmetry energy essentially characterizes the isospin dependent part of the
EOS of isospin asymmetric nuclear matter and it is defined
via the parabolic approximation to the nucleon specific energy (i.e., EOS) of
an asymmetric nuclear matter, i.e.,
\begin{equation}
E(n ,\delta)=
E_{0}(n)+E_{\rm{sym}}(n)\delta^{2}+\mathcal{O}(\delta^{4}) \, ,
\label{EOSANM}
\end{equation}
with baryon density $n=n_{\rm p}+n_{\rm n}$ and isospin asymmetry
$\delta=(n_{\rm n}-n_{\rm p})/n$ where $n_{\rm p}$ and $n_{\rm n}$ denote
the proton and neutron densities, respectively. $E_{0}(n) = E(n, \delta=0)$ is
the EOS of symmetric nuclear matter, and the symmetry energy can be expressed as
\begin{equation}
E_{\rm{sym}}(n)=\left.\frac{1}{2!}\frac{\partial^{2}E(n, \delta)}{%
\partial \delta^{2}}\right|_{\,\delta =0} \, .
\label{Esym}
\end{equation}
Because of the
exchange symmetry between protons and neutrons (isospin symmetry) in nuclear matter,
there are no odd-order $\delta$ terms in Eq. (\ref{EOSANM}).
The empirical parabolic law in Eq. (\ref{EOSANM}) for EOS of asymmetric
nuclear matter has been confirmed by all many-body theory calculations to
date, at least for densities up to moderate values~\citep{LCK08}.

Around a reference density $n_{\rm r}$, the symmetry
energy $E_{\mathrm{sym}}(n)$ can be further expanded as
\begin{equation}
E_{\mathrm{sym}}(n )=E_{\text{\textrm{sym}}}({n _{\rm r}})+\frac{L({n _{\rm r}})}{3} \left(%
\frac{n -{n _{\rm r}}}{{n _{\rm r}}}\right)+\mathcal{O}\left(\frac{n -{n _{\rm r}}}{{n _{\rm r}%
}}\right)^{2},
\end{equation}
with the density slope parameter defined as
\begin{equation}
L({n _{\rm r}})=\left.3n _{\rm r}\frac{\partial E_{\mathrm{sym}}(n )}{\partial n }\right|_{\,n
=n _{\rm r}}.  \label{L}
\end{equation}
The slope parameter $L(n_{\rm r})$ reflects the density dependence of
symmetry energy around $n_{\rm r}$.

The EOS of isospin asymmetric nuclear matter essentially determines
the structure of the conventional neutron stars (see, e.g., Ref.~\citep{XuJ09}).
In the present work, the structure of the conventional neutron stars is
obtained by solving the TOV equations, i.e.,
\begin{eqnarray}
\frac{{\rm d}M(r)}{{\rm d}r} &=& 4 \pi r^2 \epsilon(r) \\
\frac{{\rm d}P(r)}{{\rm d}r} &=&
-\frac{G \epsilon(r) M(r)}{r^2} \left[1+\frac{P(r)}{\epsilon(r)}\right] \nonumber \\
&& \times \left[1+\frac{4 \pi P(r) r^3}{M(r)}\right] \left[1-\frac{2GM(r)}{r}\right]^{-1} \, ,
\label{tov1}
\end{eqnarray}
where $G$ is Newton's gravitational constant,
$M(r)$ is the gravitational mass inside the sphere of radius
$r$, $\epsilon(r)$ and $P(r)$ are, respectively, the corresponding
total energy density and total pressure of the neutron star matter at radius $r$:
\begin{eqnarray}
\epsilon(r) &=& \epsilon_{\rm b}(r) + \epsilon_{ l}(r)
\label{endensity} \\
P(r) &=& P_{\rm b}(r) + P_{l}(r) \, .
\label{pressure01}
\end{eqnarray}
In the present paper, the natural unit $\hbar = c =1$ is adopted. The neutron star matter
is assumed to be neutrino free and only composed of neutrons, protons, electrons
and muons in $\beta$-equilibrium with charge
neutrality (i.e., the $n p e \mu$ matter). Thus the baryon part $\epsilon_{\rm b}$ of the total energy density
can be written as
\begin{equation}
\epsilon_{\rm b}(n_{\rm b},\delta_{\rm eq}) = n_{\rm b}E(n_{\rm b},\delta_{\rm eq}) + m_{\rm b}n_{\rm b} \, ,
\label{baryonden}
\end{equation}
where $m_{\rm b}$ is the baryon mass, $n_{\rm b}$ is the local baryon density at the radius $r$
and the local isospin asymmetry $\delta_{\rm eq}$
is determined by the $\beta$-equilibrium with charge neutrality condition.
The energy density of leptons $\epsilon_{l}$ is calculated using the
non-interacting Fermi gas model, and can be expressed as:
\begin{equation}
\epsilon_{l} = \eta \vartheta(\kappa) \, ,
\end{equation}
where $\eta$ and $\vartheta(\kappa)$ are defined, respectively as
\begin{eqnarray}
\eta &=& \frac{m_{l}}{8 \pi^2 \lambda^3} \nonumber \\
\vartheta(\kappa) &=& \kappa\sqrt{1+\kappa^2}(1+2\kappa^2)-\ln\left(\kappa+\sqrt{1+\kappa^2}\right) \nonumber
\end{eqnarray}
with
\begin{equation}
\lambda = \frac{1}{m_{l}} \, , \; \kappa = \lambda (3\pi^2 n_{l})^{1/3}. \nonumber
\end{equation}
In the above expressions, $m_{l}$ and $n_{l}$ are the lepton mass and density, respectively.
Then the corresponding pressure can be obtained through the thermodynamical relation:
\begin{equation}
P_{i} = n_{i}^2 \frac{{\rm d}(\epsilon_i/n_i)}{{\rm d}n_i}, \, (i = b~\text{or}~l).
\label{pressure02}
\end{equation}
For the $npe\mu$
matter, the $\beta$-equilibrium condition is
\begin{equation}
\mu_{\rm n} - \mu_{\rm p} = \mu_{\rm e} = \mu_{\mu} \, ,
\label{betaeq}
\end{equation}
where $\mu_i (i = {\rm n,p,e,\mu})$ represents the chemical
potential of the particle species $i$.
For neutrons and protons, the chemical potential is
determined by the EOS of isospin asymmetric nuclear matter via its definition as
\begin{equation}
\mu_{i} = \frac{d \big(nE(n,\delta)\big)}{d n_{i} }, \, (i = n~\text{or}~p).
\end{equation}
For leptons (electrons and muons), their chemical potentials can be expressed as
\begin{equation}
\mu_{l} = \sqrt{p^2_{ Fl} + m_{l}^2} \, ,
\label{mulepton}
\end{equation}
where the lepton's Fermi momentum is
\begin{equation}
p_{Fl} = (3 \pi^2 n_{l})^{1/3} \, .
\end{equation}
Eq.~(\ref{betaeq}) together with the charge neutrality condition
\begin{equation}
n_{\rm p} = n_{\rm e}+n_{\mu}
\label{chargeneutral}
\end{equation}
then determine the proton fraction $x_{\rm p} = n_{\rm p}/n_{\rm b}$
and the fractions of other particles as functions of
baryon density in the neutron star matter.

For the calculations of asymmetric
nuclear matter EOS, we use in the present work the standard Skyrme-Hartree-Fock (SHF) approach (see, e.g.,
Ref.~\citep{Cha97}) in which the nuclear effective
interaction is taken to have a zero-range, density- and momentum-dependent
form, i.e.,
\begin{equation}
\begin{split}
V_{12}(\mathbf{R},\mathbf{r}) = &\; t_{0} (1+x_{0}P_{\sigma})\delta(\mathbf{r}) \\
&+ \frac{1}{6}t_{3}(1+x_{3}P_{\sigma})n^{\sigma}(\mathbf{R})\delta(\mathbf{r}) \\
&+ \frac{1}{2}t_{1}(1+x_{1}P_{\sigma}) [K^{^{\prime}2}\delta(\mathbf{r})+\delta(\mathbf{r})K^{2}] \\
&+ t_{2}(1+x_{2}P_{\sigma})\mathbf{K}^{^{\prime}}\cdot \delta(\mathbf{r})\mathbf{K} \\
&+ iW_{0}(\mathbf{\sigma}_{1}+\mathbf{\sigma}_{2})\cdot [
\mathbf{K}^{^{\prime}}\times \delta(\mathbf{r})\mathbf{K} ]\, ,
\label{V12Sky}
\end{split}
\end{equation}
with $\mathbf{r}=\mathbf{r}_{1}-\mathbf{r}_{2}$ and $\mathbf{R}=(\mathbf{r}%
_{1}+\mathbf{r}_{2})/2$. In the above expression, the relative momentum
operators $\mathbf{K}=(\mathbf{\nabla}_{1}-\mathbf{\nabla}_{2})/2i$ and $%
\mathbf{K}^{\prime}=-(\mathbf{\nabla}_{1}-\mathbf{\nabla}_{2})/2i$ act on
the wave function on the right and left, respectively. The quantities $%
P_{\sigma}$ and $\sigma_{i}$ represent, respectively, the spin exchange
operator and Pauli spin matrices.

The Skyrme interaction in Eq.~(\ref{V12Sky}) includes totally $10$ parameters,
i.e., the $9$ Skyrme force parameters $\sigma$, $t_{0}-t_{3}$, $x_{0}-x_{3}$,
and the spin-orbit coupling constant $W_{0}$. This standard SHF approach has
been shown to be very successful in describing the structure of finite nuclei
as well as the properties of neutron stars~\citep{Cha97,Fri86,Klu09}.

\subsection{DM accretion onto neutron stars}

In this work, we consider DM accretion onto neutron stars by following
the basic lines of Refs.~\citep{Press85,Gould87,Chris08}.
We suppose that the neutron star seizes DM particles
from the area locating on a sphere
of radius $R_0$ centered at the centra of the neutron star.
We further let $R_0 \rightarrow \infty$, which means that $R_0$ is so large that
the gravitational field of the neutron star at $r=R_0$ is negligible.
We further assume that the neutron star is motionless
with respect to the ambient DM halo and thus a spherically symmetric accretion scenario is adopted.
Then the flux of DM crossing the
spherical surface (per unit time) with velocity ranging from
$v$ to $v+{\rm d}v$ and the angle relative to the normal line between $\alpha$
and $\alpha+{\rm d}\alpha$ is
\begin{equation}
{\rm d}F = p(v) {\rm d}v \, v\cos\alpha \, \sin\alpha \, 2\pi R_0^2 {\rm d}\alpha \, ,
\label{flux1}
\end{equation}
where $p(v)$ is the local DM velocity distribution evaluated
in the Galactic rest frame and $\alpha$ is limited within the region
$[\frac{\pi}{2}, \pi]$ to ensure that only the DM particles moving
toward the neutron star are available for accretion.
Integrating ${\rm d}F$ with respect to $v$ and $\alpha$, one can get
the maximum number of DM particles which are possible to be
captured by the neutron star.
In order to determine the integrating phase space for each variable, we
first study the trajectory for a non-relativistic DM particle traveling outside the neutron star
which is given by the so-called geodesic equation
\begin{equation}
\frac{A(r)}{r^4} \left(\frac{{\rm d}r}{{\rm d}\varphi}\right)^2 + \frac{1}{r^2} - \frac{1}{J^2 B(r)} = -\frac{e}{J^2} \, .
\label{trajec1}
\end{equation}
Here $r$ and $\varphi$ are the polar coordinates and without loss of generality, the motion of
the DM particle is taken to be in the $\theta = \pi/2$ plane.
$A(r)$ and $B(r)$ are the space-time geometry of a spherically symmetric system defined by
\begin{equation}
{\rm d}s^2 =
-B(r) {\rm d}t^2 + A(r) {\rm d}r^2 + r^2 {\rm d}\theta^2 + r^2 \sin^2 \theta {\rm d}\varphi^2 \, ,
\label{metric1}
\end{equation}
and their values are affected by the energy distribution of the whole system.
In the vacuum outside the neutron star, we have $A(r) = (1 - \frac{2GM_{\rm S}}{r})^{-1}$ and $B(r)=A^{-1}$
which are known as the Schwarzchild metrics and
are uniquely determined by the neutron star mass $M_{\rm S}$
divided by the radial distance from the center point, i.e. $M_{\rm S}/r$.
Moreover, in Eq.~(\ref{trajec1}),
the angular momentum per unit mass $J$ and the variable $e$ are defined as
\begin{eqnarray}
J &=& v R_0 \sin \alpha
\label{JAngMom}, \\
e &=& 1-2E_{0},
\label{eTotE}
\end{eqnarray}
with $E_0 = \frac{1}{2} v^2$  being the total energy per unit mass
at $r=R_0$. It should noted that both $J$ and $e$ are defined at $r=R_0$, and they are
invariant integrals of orbits both in- and outside the neutron star due to the
conservation of angular momentum and energy.
We consider the fact that the DM particles can be captured by the neutron stars only if
they are able to interact with the neutron star matter.
That is to say, the allowed variable regions for the variables in Eq.~(\ref{flux1}),
which ensure that the traveling trajectories of the DM particles pass through
the neutron star for capturing, can be obtained equivalently by requiring that the
perihelion radius (closest distance to the center of the neutron star),
$r_{\rm p}$, for each particle must be less than or equal to
the neutron star radius, $R_{\rm S}$.
The perihelion radius $r_{\rm p}$ can be obtained by letting
\begin{equation}
\left.\frac{{\rm d}r}{{\rm d}\varphi}\right|  _{r=r_{\rm p}}=0
\label{peri1}
\end{equation}
in Eq.~(\ref{trajec1}), and the requirement
$r_{\rm p}\leq R_{\rm S}$ leads to the following condition for capturing
\begin{equation}
J^2 \leq J_{\rm M}^2 = R_{\rm S}^2 \left[(1-\frac{2GM_{\rm S}}{R_{\rm S}})^{-1} - e \right] \, .
\label{jmax}
\end{equation}
So it is more convenient to express Eq.~(\ref{flux1}) in terms of $J$
and $E_0$ according to their relationship with respect to $v$ and $\alpha$ mentioned previously.
Furthermore, we assume that the DM population in the DM halo
follows a Maxwell-Boltzmann distribution of velocities in the
Standard Halo Model, that is,
\begin{equation}
p(v){\rm d}v
= n_{\chi} \left(\frac{1}{\pi v_0^2}\right)^{3/2} 4 \pi v^2 e^{-\frac{v^2}{v_0^2}} {\rm d}v \, ,
\label{maxwell}
\end{equation}
where $n_{\chi}$ is the DM number density of the DM halo around the neutron star,
and $v_0$ is the most probable speed of the velocity distribution.
Then Eq.~(\ref{flux1}) can be re-expressed as
\begin{equation}
{\rm d}F
= n_{\chi} \left(\frac{1}{\pi v_0^2}\right)^{3/2} 4 \pi^2  e^{-\frac{2 E_{0}}{v_0^2}} {\rm d}E_{0} {\rm d}J^2 \, ,
\label{flux2}
\end{equation}
and the total capture rate of the DM particles can be obtained as
\begin{equation}
\begin{split}
F &= \int_0^{\infty} \, \int_0^{J_{\rm M}^2} \, {\rm d}F \\
&=n_{\chi} \left(\frac{1}{\pi {v_0}^2}\right)^{\frac{3}{2}} 4 \pi^2 R_{\rm S}^2 \frac{{v_0}^2}{2} \left[(1-\frac{2GM_{\rm S}}{R_{\rm S}})^{-1} - 1 \right] \, ,
\label{accrate}
\end{split}
\end{equation}
where we integrate over $E_0$ from zero to infinity and
over $J^2$ from zero to $J^2_{\rm M}$ according to Eq. ({\ref{jmax}}).

\subsection{DM scattering with neutron star matter}
In order to be trapped by a neutron star, the DM particles
are required to scatter with the neutron star matter and lose
enough energy to form some bound orbits.
Since the velocities of the DM particles
approaching the surface of the neutron star are roughly equal to the
local escape velocity ($\sim 0.6 \, c$),
even one collision is enough to ensure such a bound orbit.
The fraction that the DM particles undergo
at least one collision inside the neutron star
can be expressed as~\citep{Chris08}
\begin{equation}
f =
\left\langle 1 - e^{-\int \sigma_{\chi} n_{\rm b} {\rm d}l} \right\rangle \, ,
\label{frac}
\end{equation}
where the angle brackets represent the average over all possible
DM trajectories in the above equation, and
$n_{\rm b}$ is the local baryon number density,
$\sigma_{\chi}$ denotes the effective scattering cross section
between the DM particles and nucleons inside the neutron star.

For a conventional neutron star consisting only of nucleons and leptons
as we are considering in the present work, $\sigma_{\chi}$ can be
evaluated in terms of the DM-proton and DM-neutron scattering cross sections
in infinite nuclear matter as
\begin{equation}
\begin{split}
\sigma_{\chi} &= \sigma_{\rm p} \xi_{\rm p}(r) x_{\rm p}(r) + \sigma_{\rm n} \xi_{\rm n}(r) [1-x_{\rm p}(r)]\\
              &= \sigma_{\rm p} [g_{\rm np}^2 \xi_{\rm n} + (\xi_{\rm p} - g_{\rm np}^2 \xi_{\rm n} )x_{\rm p}] \, .
\label{css}
\end{split}
\end{equation}
Here $\sigma_{\rm p,n}$ denote the DM-proton
and DM-neutron cross sections
in free space, respectively. $g_{\rm np}$ is the
so-called isospin-violating factor which is defined as
\begin{equation}
g_{\rm np} =
\frac{f_{\rm n}}{f_{\rm p}} \, ,
\label{gnp}
\end{equation}
with $f_{\rm n,p}$ denoting the effective coupling of DM to neutrons and protons, respectively.
The factor $\xi_{\rm p,n}$ takes into account the medium corrections on
the DM-nucleon scattering cross sections by considering the Pauli blocking and Fermi motion due to
the proton (neutron) degeneracy effect and it can
be analytically expressed as~\citep{Chen01}
\begin{equation}
\xi_{\rm p,n} = 1 - \frac{2}{5}\left(\frac{p_{\rm F}^{\rm p,n}}{p_{\chi}}\right)^2 \, ,
\label{Pauli01}
\end{equation}
for $p_{\chi}>p_{\rm F}^{\rm p,n}$, and it becomes
\begin{equation}
\begin{split}
\xi_{\rm p,n} = &\;  1 - \frac{2}{5}\left(\frac{p_{\rm F}^{\rm p,n}}{p_{\chi}}\right)^2\\
& + \frac{2}{5}\left(\frac{p_{\rm F}^{\rm p,n}}{p_{\chi}}\right)^2 \left[1-\left(\frac{p_{\chi}}{p_{\rm F}^{\rm p,n}}\right)^2\right]^{5/2} \, ,
\label{Pauli02}
\end{split}
\end{equation}
for $p_{\chi}<p_{\rm F}^{\rm p,n}$.
Here $p_{\chi} = m_{\chi} \frac{v_{\rm esc}}{\sqrt{1-v_{\rm esc}^2}}$ is the momentum of the incident DM particle
with $v_{\rm esc} = \sqrt{1 - B(r)}$~\citep{Lava10} denoting the local
escape velocity of the neutron star, and the Fermi momenta are given
by $p_{\rm F}^{\rm p,n} = (3 \pi^2 n_{\rm p,n})^{1/3}$.
We ignore the scattering between the DM particles and leptons
since the masses of leptons possibly existing inside
the neutron stars, e.g., $m_{\rm e}\sim 0.5 $ MeV and $m_{\mu}\sim 100$ MeV, are
much smaller than the DM mass we are interested in here ($\sim 10$ GeV),
and the scattering with these light leptons
would hardly affect the motion of the DM particles.

In Eq.~(\ref{frac}), the arc length along an orbit, ${\rm d}l$, is given by
\begin{equation}
\begin{split}
{\rm d}l &= \sqrt{A(r){\rm d}r^2 + r^2 {\rm d}\varphi^2} \\
         &= {\rm d}\varphi \frac{r^2}{\sqrt{J^2}} \sqrt{\frac{1}{B(r)}-e} \, .
\label{length}
\end{split}
\end{equation}
where we have already applied the geodesic equation (Eq.~(\ref{trajec1})) to obtain
the expression in the second line. Similarly,
we have also taken the calculation within the $\theta = \pi/2$ plane.
We would like to emphasize here that the metric in front of ${\rm d}r^2$
in Eq.~(\ref{length}) now has the following form,
that is,
\begin{equation}
A(r) = \left(1 - \frac{2GM(r)}{r}\right)^{-1},
\label{ar}
\end{equation}
where $M(r)$ is the gravitational mass within the radius $r$
instead of the mass of the whole neutron star as that in Eq.~(\ref{metric1}). Similarly,
the metric in front of the time, $B(r)$, in Eq.~(\ref{length})
is now given by
\begin{equation}
B(r) = e^{2 \phi(r)}
\label{br}
\end{equation}
with
\begin{equation}
\phi(r) = -\int_r^{\infty} \frac{G}{r'^2}\frac{M(r')+4\pi r'^3 P(r')}{1-\frac{2GM(r')}{r'}}{\rm d}r' \, .
\label{br02}
\end{equation}
Here the $\phi(r)$ corresponds to an effective Newton-like gravitational
potential (see the appendix) which is used to determine the black hole
formation conditions in the next subsection.
The integrating limit in Eq.~(\ref{br02})
is from $r$ to infinity, and one can easily verify that Eq.~(\ref{br})
returns back to the Schwarzchild-like form as that in Eq.~(\ref{metric1}) when the radius $r>R_{\rm S}$.

Finally, in Eq.~(\ref{frac}), for a given neutron star, we average over all the possible pathes of
the DM particles passing through the neutron star. Since the traveling trajectories of the DM particles
are fixed by the initial conditions of motion at $r = R_0$, Eq.~(\ref{frac})
can be further expressed according to the DM population distribution at $R_0$ as
\begin{equation}
f = \left\langle Q\right\rangle
= \frac{\int_0^{\infty} {\rm d}v \int_0^{\alpha_{\rm M}} {\rm d}\alpha \, p(v) q(\alpha)
Q}{\int_0^{\infty} {\rm d}v \int_0^{\alpha_{\rm M}} {\rm d}\alpha \, p(v) q(\alpha)} \, ,
\label{average01}
\end{equation}
where we have $Q = 1 - e^{-\int \sigma_{\chi} n_{\rm b} {\rm d}l}$,
and $p(v)$ and $q(\alpha)$ are the velocity distribution
and angular distribution of ambient DM halo, respectively. Since we are interested in the
isotropic case, the angular distribution can be reduced to $q(\alpha)=1$.
According to the discussions about Eq.~(\ref{flux1}) and Eq.~(\ref{flux2}) in the previous subsection,
it will be more convenient to express Eq.~(\ref{average01}) in terms of $J$ and $E_0$
rather than $v$ and $\alpha$.
Thus the limit of the angular integration, $\alpha_{\rm M}$, corresponds to
$J_{\rm M}$ we introduced previously according to the relation
$J_{\rm M} = v R_0 \sin \alpha_{\rm M}$ (see, Eq.~(\ref{JAngMom})),
and the integration over $E_0$ remains from $0$ to $\infty$.
Once again, by assuming that the DM population follows a Maxwell-Boltzmann distribution for
velocities as in  Eq.~(\ref{maxwell}), we can re-express Eq.~(\ref{average01}) as
\begin{eqnarray}
f &=& \frac{\int_0^{\infty} e^{-\frac{2E_{0}}{{v_0}^2}}{\rm d}E_{0}  \int_0^{J^2_{\rm M}} \left(1-\frac{J^2}{2E_{0}R_{0}^2}\right)^{-1/2} \frac{ Q(J^2,E_{0}) }{\sqrt{J^2}}{\rm d}J^2}{\int_0^{\infty} e^{-\frac{2E_{0}}{{v_0}^2}} {\rm d}E_{0} \int_0^{J^2_{\rm M}} \left(1-\frac{J^2}{2E_{0}R_{0}^2}\right)^{-1/2} \frac{1}{\sqrt{J^2}} {\rm d}J^2} \notag \\
&\approx&  \frac{\int_0^{J^2_{\rm M}} \frac{Q(J^2)}{\sqrt{J^2}} {\rm d}J^2}{2 \sqrt{J^2_{\rm M}}} \, .
\label{average02}
\end{eqnarray}
We have assumed $\left(1-\frac{J^2}{2E_{0}R_{0}^2}\right)^{-1/2}\approx 1$ and $e \approx 1$
in order to simplify the expression to the form given in the second line in Eq.~(\ref{average02}),
and these approximations are always valid for all cases satisfying $R_0 \rightarrow \infty$.

Combining Eq.~(\ref{accrate}) and Eq.~(\ref{average02}) together, we can get the total
mass $M_{\rm t}$ of the DM particles that can be trapped by a neutron star in
a period of time $t$ as
\begin{eqnarray}
M_{\rm t}
& = & 4.07 \times 10^{40} \, {\rm GeV} \, \frac{M_{\rm S} R_{\rm S}}{1 - 2.964 \frac{M_{\rm S}}{R_{\rm S}}} \left(\frac{m_\chi n_\chi}{0.3 \, {\rm GeV}/{\rm cm}^3}\right) \notag \\
& & \times \left( \frac{v_0}{220 \, {\rm km}/{\rm s}}\right)^{-1} \left(\frac{t}{{\rm Gyr}}\right) \, f \, ,
\label{taccmass}
\end{eqnarray}
where $M_{\rm S}$ is the gravitational mass of the neutron star in unit of solar mass
$M_{\odot}$ and $R_{\rm S}$ is neutron star radius in unit of km. It should be noted here that
the structure effects of neutron stars on $M_{\rm t}$ are due to the factor $f$.

We would like to point out that all the
expressions above have been obtained within the
framework of general relativity. And for Newtonian case, we note that
the trapped mass of the DM particles will be reduced
by about a factor of $2$.

\subsection{Black hole formation and neutron star destruction}

Since we focus on the non-interacting bosonic ADM in the present work,
the Chandrasekhar limit of the mass for a boson star consisting of such kind of DM
can be expressed as~\citep{Chris11,Kumar13}
\begin{equation}
M_{\rm cha} = \frac{2}{\pi} \frac{M_{\rm pl}^2}{m_{\chi}} \, ,
\label{chand}
\end{equation}
where $M_{\rm pl} = 1.22\times 10^{19} \, {\rm GeV}$ is the Planck mass.

After being captured by the neutron star, a DM particle will go on
scattering with the neutron star matter until its velocity reduces
to the thermal velocity depending on the local temperature of the neutron star matter.
The spending time of this thermalization process has been estimated by
Kouvaris in Ref.~\citep{Chris13}, and it can be parameterized as
\begin{equation}
t_{\rm th} =
0.2 \, {\rm yr} \, \left(\frac{m_{\chi}}{\rm TeV}\right)^2
\left(\frac{\sigma_{\chi}}{10^{-43}\, {\rm cm}^2}\right)^{-1}
\left(\frac{T}{10^5 \, {\rm K}}\right)^{-1} \, ,
\label{thertime}
\end{equation}
which agrees with the vast majority of results given by
other authors~\citep{Yu12,Kumar13,Ber13}. The value of
$t_{\rm th}$ is much smaller than the living age of the old neutron stars
we are interested in.
Therefore one can assume that the DM particles are totally thermalized
as soon as they are captured by the neutron star.
Thermalized DM particles will continually drift to the neutron star center
and accumulate within a typical thermal radius,
\begin{equation}
\begin{split}
r_{\rm th}
&= \left[\frac{9 T_{\rm c}}{4 \pi G  m_{\chi} B_0 (\epsilon_{\rm c}+3P_{\rm c})}\right]^{1/2} \\
&\approx 2.2 \, {\rm m} \left[\frac{\rm GeV}{m_{\chi}} \frac{T_{\rm c}}{10^5 \, {\rm K}}
\frac{\rm GeV\cdot fm^{-3}}{B_0 (\epsilon_{\rm c}+3P_{\rm c})} \right]^{1/2} \, ,
\label{therradius}
\end{split}
\end{equation}
where $B_0 = B(0)$ (see Eq.~(\ref{br})), $m_{\rm N}$ denotes the mass of nucleons, and
$T_{\rm c}$, $\epsilon_{\rm c}$ and $P_{\rm c}$ are
the temperature, energy density and pressure at the center of the neutron star, respectively.
Derivation of Eq.~(\ref{therradius}) takes into account the correction of
general relativity as detailed in the appendix.
Once the gravity of the accumulated DM particles exceeds that of the baryons
(the corresponding critical number of DM particles is $N_{\rm th}^{\rm self}= 4\pi r_{\rm th}^3 n_{\rm c}/3$
where $n_{\rm c}$ is the baryon density at the neutron star center), the self-gravitation
will take place so that the DM particles can form a boson star
at the center of the host neutron star. For $m_{\chi} \le 10^{17} \text{GeV}(T_c/10^5 K)^3$
which is always satisfied in the cases we are considering here, the boson star mass
is larger than the Chandrasekhar limit $M_{\rm cha}$ (see, e.g., Ref.~\citep{Yu12}), and thus the
gravitational collapse always occurs as soon as these thermalized DM particles become
self-gravitating in the neutron star.

In the above discussion, the thermalized DM particles have been assumed
to follow a Maxwellian velocity distribution. However, for the non-interacting bosonic DM
system with extremely high density, the formation of a macroscopic quantum state, i.e.,
BEC, confined by the neutron star's gravitational field,
can take place when the number of the thermalized DM particles exceeds (see the appendix)
\begin{equation}
\begin{split}
N_{\rm BEC}
&= \zeta(3)\left[\frac{T_{\rm c}}{\sqrt{4 \pi G B_0 (\epsilon_{\rm c}+3P_{\rm c})/3}}\right]^3 \\
&\approx 2.4 \times 10^{35} \left(\frac{T_{\rm c}}{10^5 \, {\rm K}}\right)^3
\left[\frac{\rm GeV\cdot fm^{-3}}{B_0 (\epsilon_{\rm c}+3P_{\rm c})}\right]^{\frac{3}{2}} \,
\label{BECnum}
\end{split}
\end{equation}
where $\zeta(3) = 1.202$ is the Riemann zeta function.
This condensation process happens always before
the occurrence of the self-gravitation
of the thermalized DM system
for DM roughly lighter than several TeV depending on the temperature of the neutron
star core~\citep{Yu12,Chris13} since we have $N_{\rm BEC} < N_{\rm th}^{\rm self}$.
The DM particles which exceed $N_{\rm BEC}$ will lose their kinetic energy and
go to the ground state. Being attracted by the neutron star's gravity, they will
gather within a radius (see the appendix)
\begin{equation}
\begin{split}
r_{\rm BEC} = \;&
\left[\frac{3}{8 \pi G m_{\chi}^2 B_0 (\epsilon_{\rm c}+3P_{\rm c})}\right]^{1/4} \\
\approx \;& 1.4 \times 10^{-6} \, {\rm m} \, \\
&\times \left(\frac{\rm GeV}{m_{\chi}}\right)^{1/2} \left[\frac{\rm GeV\cdot fm^{-3}}{B_0 (\epsilon_{\rm c}+3P_{\rm c})}\right]^{\frac{1}{4}}
\label{BECradius}
\end{split}
\end{equation}
which is much smaller than the thermal radius given in Eq. (\ref{therradius}).
In this case, the total mass of the DM particles in the BEC phase required to
form self-gravitating system is
\begin{equation}
\begin{split}
M_{\rm self}^{\rm BEC}
&\sim m_{\rm N} n_{\rm c} \frac{4}{3} \pi r_{\rm BEC}^3\\
&\approx 10^{28} \, {\rm GeV} \, \left(\frac{n_{\rm c}}{\rm fm^{-3}}\right)
\left(\frac{\rm GeV}{m_{\chi}}\right)^{\frac{3}{2}}
\left[\frac{\rm GeV\cdot fm^{-3}}{B_0 (\epsilon_{\rm c}+3P_{\rm c})}\right]^{\frac{3}{4}} \, .
\label{selfgrabec}
\end{split}
\end{equation}
Assuming $T_{\rm c}= 10^5 \, {\rm K}$, $n_{\rm c} \sim 1 \, {\rm fm^{-3}}$ and
$B_0 (\epsilon_{\rm c}+3P_{\rm c}) \sim 0.3 \, {\rm GeV\cdot fm^{-3}}$,
we note that $M_{\rm self}^{\rm BEC} < M_{\rm cha}$
means $m_{\chi} > 8\times 10^{-20} \, {\rm GeV}$.
Therefore, different from the case without considering BEC,
for DM with mass $m_{\chi}\sim {\rm GeV}$,
they will first form BEC state, and the DM particles in the BEC phase
then quickly become self-gravitating to form a mini boson star which then
collapses into black hole as long as the accumulated total DM mass satisfies
\begin{equation}
M_{\rm t} > M_{\rm cha} + m_{\chi} N_{\rm BEC}.
\label{bhcondition}
\end{equation}
For DM with mass $m_{\chi}\sim {\rm GeV}$ as we are considering in
the present work, the mini boson star with DM in BEC state will generally collapse
into black hole before the gravitational collapse of the boson star formed directly from thermalized
DM particles without considering BEC. Therefore, Eq.~(\ref{bhcondition})
determines the condition of black hole formation in the present work.

After black hole formation, the fate of the black hole is then determined
by the competition between the black hole accretion and the Hawking radiation.
If the black hole first evaporates through Hawking radiation,
there will be few observable consequences one can detect,
thus one can hardly place any constraints on the DM interactions from the observations of neutron stars.
In particular, the growth of the black hole is governed by the equation
\begin{eqnarray}
\frac{{\rm d} M_{\rm BH}}{{\rm d}t}
&=& \frac{4 \pi \lambda_{\rm s} m_{\rm N} n_{\rm c} (G M_{\rm BH})^2}{c_{\infty}^3}
+ \left.\frac{{\rm d} M_{\rm BH}}{{\rm d}t}\right|_{\rm DM} \notag \\
& & - \frac{1}{15360 \pi (G M_{\rm BH})^2} \, ,
\label{bhevo}
\end{eqnarray}
where $M_{\rm BH}$ is the black hole mass which equals to
the critical mass $M_{\rm cha}$ at $t = 0$.
The first term on the right hand side of Eq.~(\ref{bhevo}) denotes the accretion
rate of the neutron star matter by the black hole, and a spherically symmetric Bondi
accretion scenario is adopted for this term in the present work.
The term $\left.\frac{{\rm d} M_{\rm BH}}{{\rm d}t}\right|_{\rm DM}$ is
the accretion rate of the DM particles by the black hole,
and for DM which has already formed a BEC, its value equals to the DM capture rate by the neutron star~\citep{Kumar13}.
The last term represents the Hawking radiation which stalls the growth of the black hole inside the neutron star.

In order to specify the parameters in Eq.~(\ref{bhevo}), we
follow the derivation based on relativistic equations for the spherical
accretion onto a black hole adopted in Ref.~\citep{Sha83}.
To simplify the calculation, we first
parameterize the EOS of the neutron star matter for $n_{\rm b}>n_{\rm c}$ with a polytropic form as
\begin{equation}
P = K n_{\rm b}^{\Gamma} \, ,
\label{poly}
\end{equation}
where $P$ is the pressure of the neutron star matter, and $K$ and $\Gamma$ are constant parameters.
In this case, the non-dimensional accretion parameter $\lambda_{\rm s}$
in the first term on the right hand side of Eq.~(\ref{bhevo})
can then be given by
\begin{equation}
\lambda_{\rm s} = \frac{1}{4} \left(\frac{c_{\rm s}}{c_{\infty}}\right)^{\frac{5-3\Gamma}{\Gamma-1}}
                  (1+3 c_{\rm s}^2)^{\frac{3\Gamma -2}{2\Gamma-2}} \, ,
\label{lamdas}
\end{equation}
where $c_{\infty}$, also appearing in the first term on the right hand side of Eq.~(\ref{bhevo}), is the
neutron star matter sound speed at the center of the neutron star
which can be obtained according to its definition as
\begin{equation}
c_{\infty}^2(n_{\rm c}) \equiv \left.\frac{{\rm d}P}{{\rm d}\epsilon}\right|_{n_{\rm b} = n_{\rm c}}
\label{soundspeed}
\end{equation}
with $\epsilon$ being the energy density of the neutron star matter.
In Eq.~(\ref{lamdas}), the sound speed at the sonic point $c_{\rm s}$ corresponding
to $c_{\infty}$ is given by the relation
\begin{equation}
(1+3c_{\rm s}^2) \left(1-\frac{c_{\rm s}^2}{\Gamma - 1}\right)^2
= \left(1-\frac{c_{\infty}^2}{\Gamma - 1}\right)^2 \, .
\end{equation}
Here the sonic point is known as the ``critical point'' where the radial
velocity of the accreting flows becomes equal to the sound speed
in the neutron star matter.

As one can see from Eq.~(\ref{bhevo}), the Bondi accretion rate is proportional to the
square of the black hole mass while the Hawking radiation rate is proportional to
the inverse of the squared black hole mass. In addition,
the DM accretion rate is independent of
$M_{\rm BH}$ and can be set as a constant throughout the calculation.
This implies that the black hole will go on growing and eventually destroy the
host neutron star if its initial mass is large enough to ensure
\begin{equation}
\left.\frac{{\rm d} M_{\rm BH}}{{\rm d}t}\right|_{t = 0} > 0 \, ,
\label{condition2}
\end{equation}
otherwise it will evaporate through Hawking radiation firstly
in return.

Therefore, the fact that old neutron stars do exist in our
Galaxy indicates that we must have either
$M_{\rm t} < M_{\rm cha} + m_{\chi} N_{\rm BEC}$
or $\left.\frac{{\rm d} M_{\rm BH}}{{\rm d}t}\right|_{t = 0} < 0$
to prevent the destruction of neutron stars,
and this will put strong constraints on the scattering cross sections
between DM particles and nucleons in neutron stars.
In particular, by introducing the isospin-violating
DM in our calculations, we can thus constrain the isospin-dependent scattering
properties between this kind of DM and the nucleons inside neutron stars.

\section{Results and Discussions}
\label{sec:results}
%

\subsection{Nuclear symmetry energy effects on the structure of neutron stars}

In order to obtain the constraints on DM properties from neutron stars, we should
first figure out the structure of the host neutron star.
Specifically, the EOS of isospin asymmetric nuclear matter is a basic
ingredient to determine the properties of neutron stars as discussed earlier.
While for symmetric nuclear matter with equal fractions of neutrons
and protons, its EOS $E_0(n)$ is relatively well determined,
the EOS of asymmetric nuclear matter,
especially the density dependence of the nuclear symmetry
energy, $E_{\rm sym}(n)$, is largely unknown~\citep{LCK08}.
In particular, although the nuclear symmetry
energy at the subsaturation cross density $n_{\rm cross} \approx 0.11 \, {\rm fm^{-3}}$
is known to be around $26.65 \, {\rm MeV}$ from a recent work
by analyzing the binding energy difference of a number of heavy isotope pairs~\citep{Zhang13},
its values at other densities, especially at supra-saturation densities, are still
poorly known.

\begin{figure}[tbp!]
\includegraphics[width=7.5cm]{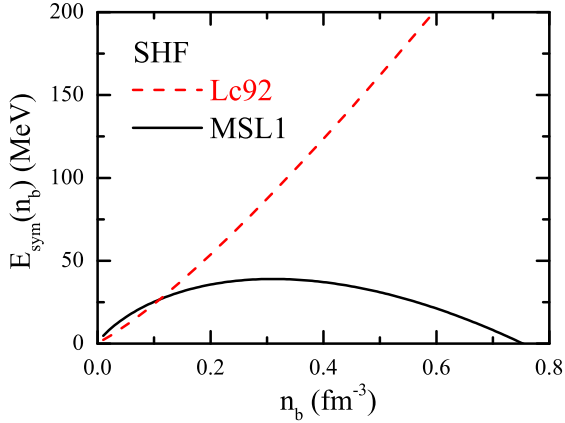}
\caption{(Color online) Density dependence
of the symmetry energy in the SHF model with MSL1 and Lc92.}
\label{esym}
\end{figure}

Shown in Fig.~{\ref{esym}} is the nuclear symmetry energy $E_{\rm sym}(n)$
as a function of baryon density in SHF calculations with
$2$ Skyrme interactions, i.e., MSL1 and Lc92~\citep{Zhang13,Zheng14}.
These two interactions are based on
the modified Skyrme-like (MSL) model~\citep{Chen10,Zhang13,Chen11a,Chen12,Chen11} in which
the $9$ Skyrme interaction parameters $\sigma $, $t_{0}-t_{3}$, $x_{0}-x_{3}$
are expressed analytically in terms of $9$ macroscopic quantities,
i.e., the saturation density $n _{0}$, $E_{0}(n_{0})$, the incompressibility $K_{0}$, the
isoscalar effective mass $m_{{\rm s},0}^{\ast }$, the isovector effective mass
$m_{{\rm v},0}^{\ast }$, $E_{\rm{sym}}({n_{\rm r}})$ at a reference density $n_{\rm r}$,
the slope parameter of the symmetry energy $L({n_{\rm r}})$, $G_{\rm S}$,
and $G_{\rm V}$. The $G_{\rm S}$ and $G_{\rm V}$ are respectively the gradient and
symmetry-gradient coefficients in the interaction part of the binding energies
for finite nuclei. Specifically, the MSL1 interaction~\citep{Zhang13} has been
obtained by fitting a number of experimental data of finite nuclei, including the binding energy,
the charge rms radius, the neutron $3p_{1/2}-3p_{3/2}$ energy level splitting in $^{208}$Pb,
isotope binding energy difference, and neutron skin data of Sn isotopes.
The Lc92 interaction has been obtained by setting
$L(n_{\rm cross}) = 92.4 \, {\rm MeV}$ in the MSL1 interaction to fit the
latest model-independent measurement of the neutron skin thickness of $^{208}$Pb
from PREX experiment at JLab~\citep{PREX12} while keeping the other $8$
macroscopic quantities and the spin-orbit coupling constant $W_{0}$ fixed at
their default values in the MSL1 interaction.
The corresponding Skyrme force parameters of these two interactions
are listed in Table~\ref{Table}.
We have selected these two interactions in such a way that they give good
descriptions for the properties of finite nuclei~\citep{Zhang13,Zheng14} while predict totally
different behaviors of $E_{\rm sym}(n)$ at supra-saturation densities as shown in Fig.~{\ref{esym}
which essentially represent the current uncertainties of the high density behaviors of the symmetry energy.

%
\begin{table}
\begin{center}
\caption{Parameters of the Skyrme forces MSL1 and Lc92.}
\begin{tabular}{lcccc}
\hline\hline
 ~~~ &~~~~~& ${\rm MSL1}$ &~~~~~& {\rm Lc92}\\
\hline
 $t_0$ $({\rm MeV\cdot fm^3})$   &  & -1963.23  &  &  -1962.93\\
 $t_1$ $({\rm MeV\cdot fm^5})$   &  & 379.845  &  &  379.795\\
 $t_2$ $({\rm MeV\cdot fm^5})$   &  & -394.554  &  &  -394.798\\
 $t_3$ $({\rm MeV\cdot fm}^{3+3\sigma})$   &  & 12174.9  &  &  12174.3\\
 $x_0$    &  & 0.32077  &  &  -0.73832\\
 $x_1$    &  & 0.344849  &  &  0.344954\\
 $x_2$    &  & -0.847304  &  &  -0.84731\\
 $x_3$    &  & 0.32193  &  &  -1.5352\\
 $\sigma$    &  & 0.26936  &  &  0.26942\\
 $W_0$ $({\rm MeV\cdot fm^5})$   &  & 113.62  &  &  113.62\\
\hline\hline
\end{tabular}
\label{Table}
\end{center}
\end{table}

\begin{figure}[tbp!]
\includegraphics[scale=0.4]{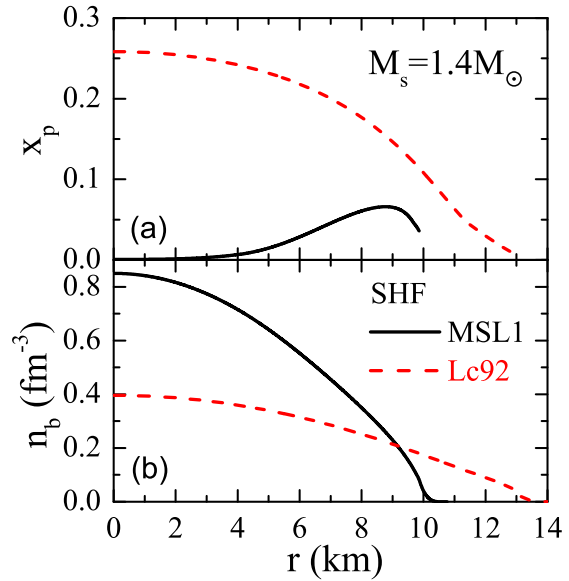}
\caption{(Color online) Proton fraction in the homogeneous liquid core (upper panel) and baryon
number density (lower panel) as functions of the radius $r$ inside
a typical neutron star with mass of $1.4 \, M_{\odot}$ from
SHF model with MSL1 (solid lines) and Lc92 (dashed lines).}
\label{composition}
\end{figure}

The global properties of static neutron stars are determined
by the EOS of neutron star matter over a broad density region ranging
from the center to the surface of neutron stars.
Generally, a typical neutron star contains the liquid core,
inner crust, and outer crust from the center to surface. For
the liquid core, we use the EOS of $n p e \mu$ matter from SHF
calculations introduced earlier. In the inner crust with densities
between $n_{\rm out}$ and the transition density $n_{\rm t}$
where the nuclear pastas may exist, because of our poor
knowledge about its EOS from first principle, following Carriere
{\it et al.}~\citep{Car03} (see also Refs.~\citep{XuJ09,Zheng12}) we construct its EOS according to
\begin{equation}
P = a + b \epsilon^{4/3} \, .
\label{innercrust}
\end{equation}
This polytropic form with an index of $4/3$ has been found
to be a good approximation to the crust EOS~\citep{Latt00,Latt01,Lat99}.
The $n_{\rm out} = 2.46\times 10^{-4} \, {\rm fm}^{-3}$
is the density separating the inner from the outer crust.
The core-crust transition density $n_{\rm t}$ of neutron stars is
determined self-consistently with the Skyrme interactions by a dynamical approach (See Refs.~\citep{XuJ09,Zheng12} for details).
The constants $a$ and $b$ are then determined by
\begin{equation}
a =
\frac{P_{\rm out} \epsilon_{\rm t}^{4/3} - P_{\rm t} \epsilon_{\rm out}^{4/3}}
{\epsilon_{\rm t}^{4/3} - \epsilon_{\rm out}^{4/3}} \, ,
\end{equation}
and
\begin{equation}
b =
\frac{P_{\rm t} - P_{\rm out}}
{\epsilon_{\rm t}^{4/3} - \epsilon_{\rm out}^{4/3}} \, ,
\end{equation}
where $P_{\rm t}$ ($\epsilon_{\rm t}$) and $P_{\rm out}$ ($\epsilon_{\rm out}$)
are the pressure (energy density) at $n_{\rm t}$ and $n_{\rm out}$, respectively.
In the outer crust with $6.93\times 10^{-13} \, {\rm fm}^{-3}  <
n < n_{\rm out}$, we use the EOS of BPS~\citep{Baym71,Iida97},
and in the region of $4.73\times 10^{-15} \, {\rm fm}^{-3}  <
n < 6.93\times 10^{-13} \, {\rm fm}^{-3}$
we use the EOS of Feynman-Metropolis-Teller~\citep{Baym71}.

Using the EOS constructed above, one can solve the
TOV equations (Eq.~(\ref{tov1})) to obtain the global properties
of static neutron stars. In the upper panel of Fig.~\ref{composition}, we show
the proton fraction in the homogeneous liquid core
inside the neutron star as a function of
the radius $r$ for a typical neutron star with mass of $1.4 \, M_{\odot}$
by using MSL1 and Lc92. It is interesting to see that different Skyrme interactions predict
totally different proton distributions inside the liquid core of the neutron star.
In particular, for the MSL1 interaction which predicts
a very soft symmetry energy as shown in Fig.~\ref{esym},
it is suggested that the neutron star core is almost comprised
of pure neutron matter. This feature can be understood from the fact
that the symmetry energy predicted by MSL1 reduces to zero or even becomes
negative at high density region (see Fig.~\ref{esym}) which will lead to
small or even vanishing proton fraction due to $\beta$-stability and charge
neutrality conditions (see, e.g., Ref.~{\citep{XuJ09}} and Section~\ref{subsec:EsymNStar}).
On the other hand, for Lc92 which predicts increasing relationship between
$E_{\rm sym}(n)$ and $n$ as shown in Fig.~\ref{esym},
large proton fraction is obtained in the neutron star core.
Therefore, these results indicate that the neutron and proton compositions inside
neutron stars depend strongly on the density dependence of the symmetry energy.

Furthermore, we show the corresponding radial distributions of
the baryon number density
in the lower panel of Fig.~\ref{composition}. Different from the cases of $x_{\rm p}$,
the shapes of the curves are qualitatively
the same for these two interactions.
However, a harder symmetry energy (i.e., Lc92) tends to predict
a larger radius of the neutron star and correspondingly a smaller central density
in order to keep the neutron star mass remaining the same.
In addition, we note that the pressure predicted by different
Skyrme interactions exhibits similar radial distribution as the corresponding
$n_{\rm b}(r)$ inside the neutron star. It should be mentioned that $x_{\rm p}$,
$n_{\rm b}$ and the pressure $P$ all are needed to calculate
the ``efficiency'' factor $f$ in Eq.~(\ref{frac}), and their variations
shown in Fig.~\ref{composition} could lead to significant
symmetry energy effects on the constraints on the scattering
cross sections between DM and nucleons which we will detail below.

\subsection{Effects of isospin-violating interaction
and the symmetry energy on the $\sigma_{\rm p}$ bounds from old neutron stars}
\label{sub:bounds}

\begin{figure}[tbp!]
\includegraphics[width=8.5cm]{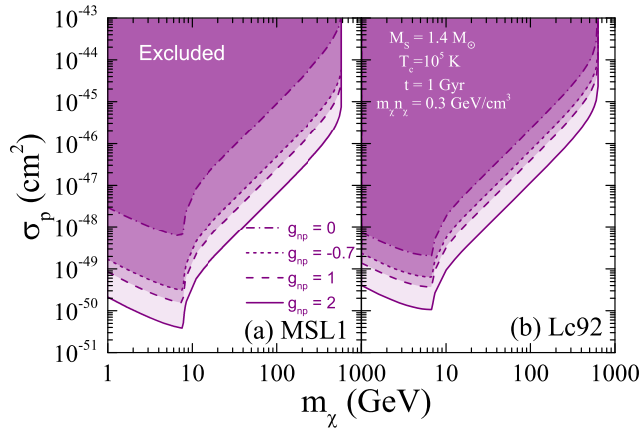}
\caption{(Color online) Exclusion contour in $m_{\chi}$-$\sigma_{\rm p}$ plane for
the non-interacting bosonic asymmetric IVDM
from a typical neutron star with $M_{\rm S} = 1.4 M_{\odot}$, $T_c = 10^5$ K, $t = 1 \, {\rm Gyr}$
and $m_{\chi} n_{\chi}=0.3 \, {\rm GeV}/{\rm cm}^3$ within SHF model with MSL1 (a) and Lc92 (b) for $g_{\rm np} = 0$, $-0.7$,
$1$ and $2$.}
\label{css01}
\end{figure}
For the first time, owing to the consideration of the realistic neutron star model,
we are able to investigate the constraint on $\sigma_{\rm p}$ for non-interacting
bosonic asymmetric IVDM from the existence of old neutron stars. In this subsection, 
we focus on the effects of isospin-violating interaction and the symmetry energy.
If the captured DM particles do form a BEC inside the neutron star,
the bounds on $\sigma_{\rm p}$ can be obtained by requesting Eq.~(\ref{bhcondition})
and Eq.~(\ref{condition2}) to be satisfied simultaneously.
Since both IVDM and ADM are typically modeled in the GeV mass region,
in this work, we restrict our interests in DM with mass ranging
from $1 \, {\rm GeV}$ to $1 \, {\rm TeV}$ where a number of
discoveries and constraints have been claimed or placed by the direct detection experiments.
We first list in Table~\ref{Table2} the values of the parameters $\Gamma$,
$c_{\infty}$, $c_{\rm s}$ and $\lambda_{\rm s}$ for MSL1 and Lc92, which are needed to
calculate the black hole evolution.

%
\begin{table}
\begin{center}
\caption{Quantities used for black hole accretion for MSL1 and Lc92.}
\begin{tabular}{lcccc}
\hline\hline
 ~~~ &~~~~~& ${\rm MSL1}$ &~~~~~& {\rm Lc92}\\
\hline
 $\Gamma$   &  & 2.546  &  &  2.600\\
 $c_{\infty}$   &  & 0.634  &  &  0.517\\
 $c_{\rm s}$   &  & 0.983  &  &  0.954\\
 $\lambda_{\rm s}$   &  & 1.415  &  &  0.929\\
\hline\hline
\end{tabular}
\label{Table2}
\end{center}
\end{table}

Shown in Fig.~\ref{css01} is the exclusion contour in the $m_{\chi}$-$\sigma_{\rm p}$
plane obtained by using the fiducial neutron star parameters with} a mass of $M_{\rm S} = 1.4 M_{\odot}$, a living age
$t = 1 \, {\rm Gyr}$ and core temperature $T_{\rm c} = 10^5 \, {\rm K}$~\citep{Latt04,Chris08,Gon10}.
Neutron stars with heavier mass and longer age
can accrete more DM particles, thus more stringent constraints will be set through Eq.~(\ref{taccmass}).
On the other hand, the higher is the core temperature,
the more DM particles are required for gravitational collapse (see Eqs.~(\ref{BECnum}) and (\ref{bhcondition})).
In particular, since one has $N_{\rm BEC}\sim T_{\rm c}^3$ according
to Eq.~(\ref{bhcondition}), the constraints will be weakened drastically when the core temperature becomes higher.
For neutron stars which are not far from the earth, we use the standard
astrophysical parameters in the Standard Halo Model for DM halo, namely, a Maxwell-Boltzmann
distribution for $p(v)$ with $v_0=220 \, {\rm km}/{\rm s}$ and a DM mass density of
$m_\chi n_{\chi}=0.3 \, {\rm GeV}/{\rm cm}^3$~\citep{Smith07} in our calculations.
Since the scattering process happening inside the neutron star is
independent of the DM halo properties and the capture
rate depends on $n_{\chi}$ and $v_0$ simply through the factor $n_{\chi}/v_0$
(see Eq.~(\ref{taccmass})), variation of the DM halo properties just rescales the
bound on $\sigma_{\rm p}$ according to $n_{\chi}/v_0$ directly.
Additionally, although DM should be, in principle, trapped
using a combination of scattering in both the core and the
crust of the neutron star, for DM first approaching to
the neutron stars with high energy, it can scatter efficiently in the core
but inefficiently in the crust due to the low matter density as well as the
coherent DM-nucleus scattering inside the crust~\citep{Horo12}.
Therefore, in Fig.~\ref{css01}, we have ignored the scattering
occurred inside the neutron star crust when we calculate the capture rate from
Eq.~(\ref{frac}) and Eq.~(\ref{taccmass}).

It should be noted that in Fig.~\ref{css01}, there exists a
bending point around several GeV mass region at which the Hawking radiation rate 
starts to dominate the inequalities
Eqs.~(\ref{bhcondition}) and~(\ref{condition2}).
That is, while bounds are obtained from the black hole formation
condition presented by Eq.~(\ref{bhcondition}) for lighter DM particles (on the
left of the bending point),
for heavier DM (on the right of the bending point),
a larger $\sigma_{\rm p}$ (i.e., a larger capture rate of DM) is required to
ensure the growth of the black hole
according to the evolution function Eq.~(\ref{bhevo}).
In addition, the cut-off mass lines around several hundred GeV in Fig.~\ref{css01} imply that
the neutron star is unable to constrain DM heavier than the cut-off mass
since the Hawking radiation rate will be always stronger than the accretion rate in Eq.~(\ref{bhevo})
for these heavy DM particles.
It is interesting to see that, by taking into account the more realistic
EOSs in the calculation for the black hole
evolution, the location of the bending point (i.e., $7.43$ GeV for MSL1 and $6.44$
GeV for Lc92) reduces significantly compared with the values obtained by
some other researchers, e.g. $16 \, {\rm GeV}$ in Ref.~\citep{Chris11}
and $13 \, {\rm GeV}$ in Ref.~\citep{Yu12}. Our results thus suggest that the location 
of the bending point depends significantly on the EOS of the neutron star matter.

Moreover, one can see from Fig.~\ref{css01} that the introduction
of isospin-violating interaction could significantly affect the bounds on the
$\sigma_{\rm p}$. For instance, in the case of
$g_{\rm np} = 0$, which means that DM will only scatter with the protons inside
the neutron star, bounds on $\sigma_{\rm p}$ would be weakened by more than
an order of magnitude relative to the normal case with $g_{\rm np} = 1$.
On the other hand, a larger $g_{\rm np}$ (e.g., $g_{\rm np}=2$) could in turn
strengthen the bounds apparently.
Bounds on $\sigma_{\rm p}$ for IVDM with $g_{\rm np} = -0.7$ are also included in Fig.~\ref{css01}.
This $g_{\rm np}$ value was firstly suggested by Feng {\it et al.}~\citep{Feng11},
and it leads to nearly complete destructive interference of the scattering
amplitudes for DM-proton and DM-neutron collisions for xenon-based detectors in
the terrestrial direct detection experiments, and has been successfully applied
to ameliorate the tension between CDMS-II(Si) and other experiments such
as LUX and SuperCDMS (see, e.g., Ref.~\citep{Zheng14}). Due to the incoherent DM-nucleon scattering inside the
liquid core of neutron stars, the effective scattering cross section
is uniquely determined by the absolute value of $g_{\rm np}$ (see Eq.~(\ref{css})),
thus the upper limit bounds with $g_{\rm np} = -0.7$ are set between the
bounds calculated with $g_{\rm np} = 0$ and $g_{\rm np} = 1$.

In addition, it is seen from Fig.~\ref{css01} that
the weakening or strengthening amplitudes for the bound on
$\sigma_{\rm p}$ strongly depend on the nuclear interactions
used. In particular, for the MSL1 interaction, which predicts a soft
symmetry energy and an absence of protons in
the neutron star core, the isospin-violating effects are much more
significant than that in the Lc92 case with a stiffer symmetry energy
and more protons inside the neutron star core.

\begin{figure}[tbp!]
\includegraphics[width=8.5cm]{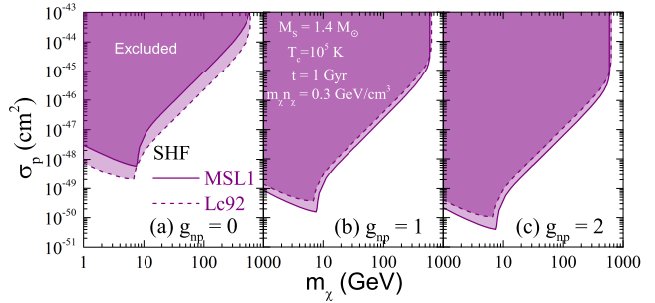}
\caption{(Color online) Exclusion contour in $m_{\chi}$-$\sigma_{\rm p}$ plane for
the non-interacting bosonic asymmetric IVDM from a typical neutron star
with $M_{\rm S} = 1.4 M_{\odot}$, $T_c = 10^5$ K, $t = 1 \, {\rm Gyr}$ and
$m_{\chi} n_{\chi}=0.3 \, {\rm GeV}/{\rm cm}^3$ within SHF model with MSL1 and
Lc92 for $g_{\rm np} = 0$ (a), $1$ (b) and $2$ (c).}
\label{css02}
\end{figure}

In order to see more clearly the symmetry energy effects on the bound
on $\sigma_{\rm p}$, we show in Fig.~\ref{css02} the exclusion contour
of $\sigma_{\rm p}$ with MSL1 and Lc92 in the same panel for different
values of $g_{\rm np}$. Indeed, one can clearly see from Fig.~\ref{css02} that
the bound obtained from the MSL1 interaction with a soft
symmetry energy is different apparently from that obtained
from the Lc92 interaction with a stiffer symmetry energy.
However, the specific values of these bounds vary with the isospin-violating factor $g_{\rm np}$
obviously. That is, on one hand, the MSL1 interaction predicts
a neutron star with smaller radius but denser matter distribution relative to
that predicted by the Lc92 interaction
(see the lower panel of Fig.~\ref{composition}).
In this case, the factor $f$ in Eq.~(\ref{frac}) obtained from MSL1 will
be larger than that in the Lc92 case, leading to a larger value of $M_{\rm t}$
as well as a smaller value of $\sigma_{\rm p}$ (panels (b) and (c) of Fig.~\ref{css02}).
On the other hand, since the proton distribution predicted by each interaction
is much different from each other, when the absolute value
of $g_{\rm np}$ decreases (which means that the scattering process becomes more sensitive to
the proton distribution inside the star),
neutron stars with fewer protons (MSL1 case) become more
transparent to the approaching DM particles, and thus the bound on $\sigma_{\rm p}$
weakens drastically in this case (panel (a) of Fig.~\ref{css02}).

Moreover, both the bending points and the cut-off mass lines obtained from these $2$ interactions
are also different from each other in Fig.~\ref{css02}. These variations further increase the 
differences among different bounds and lead to the overlap of various curves in Fig.~\ref{css02}.

\subsection{The $\sigma_{\rm p}$ bounds from observation of realistic neutron stars}
\label{sub:bounds02}

Based on the above discussions, in this subsection,
we present the results of the constraint on $\sigma_{\rm p}$ for the non-interacting bosonic asymmetric IVDM
from the observation of realistic old neutron stars. To avoid the complexity due to
the evolution history of neutron stars in binary system, here we study an isolated old
neutron star, i.e., PSR B1257+12~\citep{Wo90}. The pulsar
PSR B1257+12 is a planetary system with one solitary neutron star being orbited
by three planets locating $0.6 \, {\rm kpc}$ away from the solar system~\citep{Wo92,Wo94}.
When the neutron star's spin period and period derivative are accounted for,
its age is determined to be $0.862 \, {\rm Gyr}$~\citep{Man05}.

\begin{figure*}[tbp!]
\centering
\includegraphics[width=14.5cm]{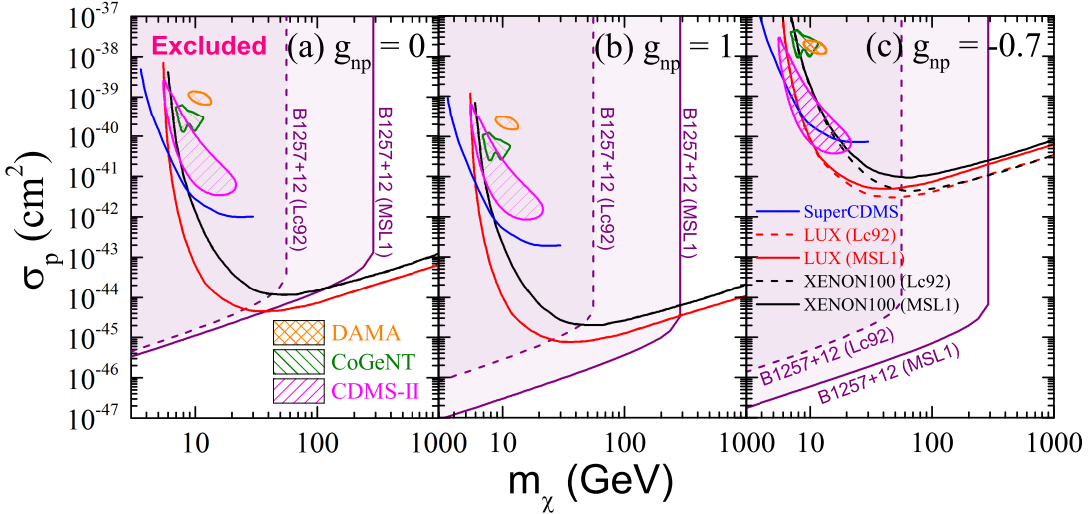}
\caption{(Color online) Exclusion contour in $m_{\chi}$-$\sigma_{\rm p}$ plane for
the non-interacting bosonic asymmetric IVDM from PSR B1257+12 within SHF model with
MSL1 and Lc92 for $g_{\rm np} = 0$ (a), $1$ (b) and $-0.7$ (c). The corresponding
results from direct detection experiments, i.e., DAMA, CoGeNT, CDMS-II(Si), XENON100, LUX
and SuperCDMS(Ge), are also included for comparison.}
\label{B1257}
\end{figure*}

For isolated neutron stars, their internal temperature is largely uncertain
in observations, and usually one can use a simple
analytical approximation to determine their
internal temperature from the well studied surface temperature
($T_{\rm s}$)~\citep{Gud82}, i.e.,
\begin{equation}
T_{\rm b} = 1.288\times10^8 \, {\rm K} \left[\frac{10^{14} \, {\rm cm/s^2}}{g_{\rm s}}
\left(\frac{T_{\rm s}}{10^6 \, {\rm K}}\right)^4\right]^{0.455} \,
\label{Tc}
\end{equation}
where $g_{\rm s} = GM_{\rm S}e^{-\phi(R_{\rm S})}/R_{\rm S}^2$ is the surface gravity and $T_{\rm b}$
is the temperature at the bottom of the heat-blanketing envelope of the neutron star.
Since the interior of a neutron star becomes isothermal within a few years after its birth~\citep{Pa04},
we assume that the core temperature $T_{\rm c}$ of the neutron star equals to
$T_{\rm b}$ in Eq.~(\ref{Tc}) in this work.
Theoretically, the time evolution of the surface temperature
of isolated neutron stars has been relatively well established
and for solitary neutron stars older than several million years,
their surface temperature is generally expected to be lower than
$10^5 \, {\rm K}$~\citep{Ya04}.
In the present work, we assume a fixed value of
$T_{\rm s} = 10^5 \, {\rm K}$ for PSR B1257+12, and the $\sigma_{\rm p}$
bounds obtained from PSR B1257+12 with this assumption thus represent a conservative constraint since a lower $T_{\rm s}$ will lead to a stronger $\sigma_{\rm p}$ bound as will be shown later.

Adopting the standard neutron star mass of $1.4 \, M_{\odot}$ for PSR B1257+12~\citep{Wo92},
we show in Fig.~\ref{B1257}
the bounds for $\sigma_{\rm p}$ from the existence of the neutron star PSR B1257+12
within SHF model with MSL1 and
Lc92 for $g_{\rm np} = 0$, $1$, and $-0.7$, respectively.
In the calculations, the halo parameters around the nearby neutron star PSR B1257+12
are assumed to have the same values as that around the earth, i.e., a Maxwell-Boltzmann
distribution for $p(v)$ with $v_0=220 \, {\rm km}/{\rm s}$ and a DM density of
$m_{\chi}n_{\chi}=0.3 \, {\rm GeV}/{\rm cm}^3$.
For comparison, we also show in Fig.~\ref{B1257} the corresponding results of the limits and
regions from the direct detection experiments including
the $90 \%$ confidence level (C.L.) limits from
XENON100~\citep{XENON100v12}, LUX~\citep{LUX13} and SuperCDMS(Ge)~\citep{SCDMS14}
along with the $90 \%$ C.L. favored regions from DAMA~\citep{DAMA09},
CoGeNT~\citep{CoGeNT11} and CDMS-II(Si)~\citep{CDMS13v1}.
The results of the direct
detection experiments for various $g_{\rm np}$ are obtained
according to the way introduced in Ref.~\citep{Zheng14}.

One can see from Fig.~\ref{B1257} that the
bounds set by PSR B1257+12 are much stronger than that given by the direct
detection experiments in the mass region about $10 \, {\rm GeV}$ where a number of
DM signals have been reported.
Especially, it is interesting to see in Fig.~\ref{B1257}~(c) that,
for $g_{\rm np} = -0.7$,
which corresponds to the so-called xenophobic IVDM~\citep{Feng13a}, while the
tension among various direct detection experiments is largely
ameliorated due to the destructive interference of DM scattering
with protons and neutrons inside the target nuclei (especially for Xenon target; see, e.g.,~\citep{Zheng14}),
all the favored DM regions reported in direct detection experiments (i.e., DAMA,
CoGeNT, and CDMS-II(Si)) are excluded by more than $3$
orders of magnitude by the new limits obtained in the present work from the
existence of PSR B1257+12.

In addition, comparing with the results shown in Fig.~\ref{css02},
one can see from Fig.~\ref{B1257} that the exclusion contours given by PSR B1257+12
become much weaker and the symmetry energy effects on the upper limits on $\sigma_{\rm p}$
as well as the cut-off mass lines become more pronounced.
In particular, the values of the upper limits of $\sigma_{\rm p}$ are
increased significantly and become comparable with
(or even larger than) the limits given by LUX and XENON100 for massive IVDM
in Fig.~\ref{B1257}~(a) and~(b).
Such a weakening effect of the constraint on $\sigma_{\rm p}$ given by PSR B1257+12 is mainly due to the
high surface temperature $T_{\rm s} = 10^5 \, {\rm K}$ used for PSR B1257+12
which leads to a much higher core temperature
$T_{\rm c} = 1.95\times10^6 \, {\rm K} \left(\frac{10^{14} \, {\rm cm/s^2}}{g_{\rm s}}\right)^{0.455}$
(i.e., $1.4 \times 10^6 \, {\rm K}$ for MSL1 and $1.9 \times 10^6 \, {\rm K}$ for Lc92)
than the fiducial value of $10^5 \, {\rm K}$ used in Fig.~\ref{css01} and Fig.~\ref{css02}.
As a result, the value of $N_{\rm BEC}$ in Eq.~(\ref{BECnum}) increases drastically,
which dominates the inequality Eq.~(\ref{bhcondition}) and
weakens the bounds on $\sigma_{\rm p}$ by more than two orders of magnitude.

\section{Conclusions and outlook}
\label{sec:conclusion}
For non-interacting bosonic asymmetric dark matter (DM), the absence of the Fermi pressure
and the ignorable dark matter self-annihilation could cause the formation
of a DM boson star in the center of neutron stars through the interactions
between DM particles and the neutron star matter, and the boson star could
reach the Chandrasekhar limit to collapse into a black hole
which may eventually destroy the host neutron star. The existence of old
neutron stars therefore can provide important constraints on the
interactions between the DM particles and nucleons.
Since the nuclear matter inside old neutron stars is generally highly
isospin asymmetric (i.e., extremely neutron-rich), it is thus specially
interesting to study the neutron star probe for isospin-violating dark matter (IVDM)
which couples differently with protons and neutrons and provides a
promising mechanism to ameliorate the tension among recent direct detection
experiments.

In the present work, for the first time, by considering a realistic neutron star model in
which the neutron star is assumed to be composed of $\beta$-stable and
electrically neutral $n p e \mu$ matter and its structure is obtained
by solving the Tolman-Oppenheimer-Volkoff equations with equation of state
of isospin asymmetric nuclear matter constrained by terrestrial experiments,
we have investigated how the isospin violating DM-nucleon interactions and
the symmetry energy in equation of state of isospin asymmetric nuclear
matter influence the extraction of the bound on the DM-proton scattering
cross-section $\sigma_{\rm p}$ from the existence of old neutron stars.
In particular, we have restricted our attention on the non-interacting
bosonic asymmetric IVDM within the mass region from
$1 \, {\rm GeV}$ to $1 \, {\rm TeV}$
which is being extensively investigated by DM direct detection experiments.
In addition, considering the large uncertainty on the density dependence
of the symmetry energy, we have employed two EOSs with very
different density dependences for the symmetry energy within the standard
Skyrme-Hartree-Fock mean field model. Furthermore, all results given in
this work, including the DM accretion by neutron stars and the BEC formation
in the center of neutron stars, have been obtained self-consistently within
the framework of general relativity.

Our results have indicated that, for a typical neutron star with mass
$M_{\rm S} = 1.4 M_{\odot}$, center temperature $T_c = 10^5$ K, living age
$t = 1 \, {\rm Gyr}$ and DM halo density $m_{\chi} n_{\chi}=0.3 \, {\rm GeV}/{\rm cm}^3$, the
bounds on the $\sigma_{\rm p}$ can be varied by more than an order of
magnitude depending on the specific values of the DM neutron-to-proton
coupling ratio $f_{\rm n}/f_{\rm p}$ we adopted. Our results have also
indicated that by considering the more realistic neutron star model rather
than a simple uniform neutron sphere as usual, the symmetry energy effects
which are presently largely unknown can also change the extraction of the
bound on $\sigma_{\rm p}$ by more than a factor of $2$.

Furthermore, we have studied how the observation of the realistic neutron
star PSR B1257+12 constrains the $\sigma_{\rm p}$. Our results have indicated that
the observed nearby isolated old neutron star PSR B1257+12 can set a stringent limit
for low-mass DM particles  ($ \le 20 \, {\rm GeV}$) that reaches a sensitivity
beyond the current best limits from direct detection experiments and excludes
the DM interpretation of all previously-reported positive direct detection experimental
results. Especially, our results have demonstrated that the existence of PSR B1257+12
excludes all the claimed DM regions reported in direct detection experiments (i.e., DAMA,
CoGeNT, and CDMS-II(Si)) by more than $3$ orders of magnitude for the specific
IVDM with $g_{\rm np} = -0.7$, namely, the so-called xenophobic IVDM, which has
been proposed to ameliorate the tension among various direct
detection experiments.

Our results in the present work are based on the assumption that IVDM is
non-interacting bosonic asymmetric dark matter. It will be interesting to see
how our results change if IVDM is self-interacting bosonic/fermionic asymmetric
dark matter. These studies are in progress and will be reported elsewhere.

\begin{acknowledgments}
This work was supported in part by the Major State Basic
Research Development Program (973 Program) in China under Contract No. 2015CB856904,
the NNSF of China under Grant Nos. 11275125 and 11135011, the ``Shu Guang" project
supported by Shanghai Municipal Education Commission and Shanghai Education Development
Foundation, the Program for Professor of Special Appointment (Eastern Scholar)
at Shanghai Institutions of Higher Learning, and the Science and Technology
Commission of Shanghai Municipality (11DZ2260700).
\end{acknowledgments}


\appendix

\section{DM thermalization and BEC formation in neutron stars}
\label{sec:appendix}
It is important to take the general relativity corrections into account
when we study the DM thermalization and BEC transition in neutron stars.
Due to the strong gravitational field of the host neutron star,
the space-time geometry in the neutron star core is warped so heavily that
all results derived in Minkowski geometry should be in principle re-examined.
Several attempts have been made recently~\citep{Jami13},
and we give a more precise derivation in this appendix.

The particle motion is governed by the geodesic equations.
For DM particles with non-zero mass, by adopting the space-time geometry in Eq.~(\ref{metric1}),
the geodesic equations can be expressed as (within the $\theta = \pi/2$ plane)
\begin{eqnarray}
0 &=& \frac{{\rm d}^2 r}{{\rm d} \tau^2} + \frac{A^{'}}{2 A} \left(\frac{{\rm d} r}{{\rm d} \tau} \right)^2 \notag \\
  &&-\frac{r}{A} \left(\frac{{\rm d} \varphi}{{\rm d} \tau} \right)^2
+\frac{B^{'}}{2 A} \left(\frac{{\rm d} t}{{\rm d} \tau} \right)^2 \label{geodesic01} \\
0 &=& \frac{{\rm d}^2 \varphi}{{\rm d} \tau^2} + \frac{2}{r} \frac{{\rm d} \varphi}{{\rm d} \tau} \frac{{\rm d} r}{{\rm d} \tau} \label{geodesic02}\\
0 &=& \frac{{\rm d}^2 t}{{\rm d} \tau^2} + \frac{B^{'}}{B} \frac{{\rm d} t}{{\rm d} \tau} \frac{{\rm d} r}{{\rm d} \tau} \, ,
\label{geodesic03}
\end{eqnarray}
where $\tau$ is the proper time and the prime denotes ${\rm d}/{\rm d}r$.
The metric components $A$ and $B$ are given by Eqs.~(\ref{ar}) and (\ref{br}), respectively.
Especially, since the energy density $\epsilon(r)$ and pressure $P(r)$
around the neutron star center are approximately uniformly distributed, we treat
them as constants (i.e., $\epsilon_{\rm c}$ and $P_{\rm c}$) in this work.
Thus the gravitational mass $M(r)$ can be expressed as
$M(r) = 4 \pi r^3 \epsilon_{\rm c} /3$.
Using the typical values of $\epsilon_{\rm c} \sim 1 \, {\rm GeV}$
and $r \sim 1 \, {\rm m}$, we then have $2G M(r)/r \sim 10^{-8} \ll 1$ which means that
the spatial part of the metric is almost Minkowski-like ($A(r) = 1$ according to Eq.~(\ref{ar})),
and the geodesic equations can be further simplified as
\begin{eqnarray}
0 &=& \frac{{\rm d}^2 r}{{\rm d} \tau^2} - r \left(\frac{{\rm d} \varphi}{{\rm d} \tau} \right)^2
+ \frac{B^{'}}{2} \left(\frac{{\rm d} t}{{\rm d} \tau} \right)^2 \label{geodesic04}\\
0 &=& \frac{{\rm d}^2 \varphi}{{\rm d} \tau^2} + \frac{2}{r} \frac{{\rm d} \varphi}{{\rm d} \tau} \frac{{\rm d} r}{{\rm d} \tau} \label{geodesic05}\\
0 &=& \frac{{\rm d}^2 t}{{\rm d} \tau^2} + \frac{B^{'}}{B} \frac{{\rm d} t}{{\rm d} \tau} \frac{{\rm d} r}{{\rm d} \tau} \, .
\label{geodesic06}
\end{eqnarray}
Under the low speed approximation of $|{\rm d} \mathbf{x}/{\rm d}t| \ll 1$,
where $\mathbf{x}$ denotes the spatial part of the space-time coordinate of DM,
Eqs. (\ref{geodesic04}) to (\ref{geodesic06}) can be rewritten as
\begin{equation}
\frac{{\rm d}^2 \mathbf{x}}{{\rm d} t^2} = - \nabla \Phi(r) \, .
\label{motion01}
\end{equation}
Here we have introduced an effective potential $\Phi(r)$ which
is defined by
\begin{equation}
\Phi^{'}(r) = B^{'}(r)/2\, .
\label{potential}
\end{equation}
Substituting Eqs.~(\ref{br}) and~(\ref{br02}) into the above equation, we then obtain
\begin{equation}
\begin{split}
\Phi^{'}(r) &= B(r) \phi^{'}(r) \\
&\simeq \frac{4 \pi}{3} G r B_0 (\epsilon_{\rm c} + 3P_{\rm c}) \, .
\label{potential02}
\end{split}
\end{equation}
The above equation is just the same as that of the Newton's gravitational potential
except for replacing the mass density by $B_0 (\epsilon_{\rm c} + 3P_{\rm c})$.
Here $B_{0} = B(0)$ is the correction of the potential due to the non-trivial space-time curvature
at the neutron star center (we note that $B_{0}$ has been simply set to be $1$ in Ref.~\citep{Jami13}).
Therefore, the motion of a DM particle in the central region of
the neutron star can be treated approximately as that it is moving in an external
potential $\Phi(r)$.
For thermalized DM following the Maxwell-Boltzmann distribution~\citep{Chris11,Yu12,Kumar13}
\begin{equation}
p_{\rm th}(v) = \sqrt{\frac{2}{\pi}\left(\frac{m_\chi}{T_{\rm c}}\right)^3}\, v^2 \exp \left(\frac{-m_\chi v^2}{2T_{\rm c}}\right) \, ,
\label{maxwell02}
\end{equation}
the typical thermal radius $r_{\rm th}$ given in Eq.~(\ref{therradius}) can be obtained
by assuming the average kinetic energy of the distribution equals to the potential
energy, i.e.,
\begin{equation}
m_{\chi} \Phi(r_{\rm th}) = \frac{1}{2}m_{\chi} v^2_{\rm rms} \, ,
\end{equation}
where $v_{\rm rms} = \sqrt{3 T_{\rm c}/m_{\chi}}$ is the root mean square speed of the distribution.
The similar process can be applied to derive Eq.~(\ref{BECradius})
for the BEC radius by assuming the zero point energy of the DM
in the BEC phase confined in the potential $\Phi(r)$ equals to the potential energy.

Moreover, the BEC in an external potential has been well studied in Ref.~\citep{Bag87}.
For a fixed temperature $T_{\rm c}$, the critical number $N_{\rm BEC}$ of scalar particles
with mass $m$ in a harmonic oscillator potential $U(r) = \varepsilon (r/a)^2$ is given by
\begin{equation}
N_{\rm BEC} = \zeta(3) T_{\rm c}^3 \left(\frac{m a^2}{2 \varepsilon}\right)^{\frac{3}{2}} \, .
\label{becnum02}
\end{equation}
Using the potential $\Phi(r)$ derived above, we then obtain
the critical number in Eq.~(\ref{BECnum}).

%


\begin{thebibliography}{99}

\bibitem[Aalseth et al. (2011)] {CoGeNT11}
Aalseth, C. E., et al. 2011,
Phys. Rev. Lett., 106, 131301

\bibitem[Abrahamyan et al. (2012)] {PREX12}
Abrahamyan, S., et al. 2012,
Phys. Rev. Lett., 108, 112502

\bibitem[Ade et al. (2013)] {Plk13}
Ade, P. A. R., et al. 2013,
arXiv:1303.5076

\bibitem[Agnese et al. (2013)] {CDMS13v1}
Agnese, R., et al. 2013,
Phys. Rev. D, 88, 031104

\bibitem[Agnese \& Ahmed et al. (2013)] {CDMS13v2}
Agnese, R., et al. 2013,
Phys. Rev. Lett., 111, 251301

\bibitem[Agnese et al. (2014)] {SCDMS14}
Agnese, R., et al. 2014,
arXiv:1402.7137

\bibitem[Akerib et al. (2014)] {LUX13}
Akerib, D. S., et al. 2014,
Phys. Rev. Lett., 112, 091303

\bibitem[Angloher et al. (2012)] {CRESST12}
Angloher, G., et al. 2012,
European Physical Journal  C, 72, 1971

\bibitem[Aprile et al. (2011)] {XENON100v11}
Aprile, E., et al. 2011,
Phys. Rev. Lett., 107, 131302

\bibitem[Aprile et al. (2012)] {XENON100v12}
Aprile, E., et al. 2012,
Phys. Rev. Lett., 109, 181301





\bibitem[Bagnato et al. (1987)] {Bag87}
Bagnato, V., Pritchard, D. E., \& Kleppner, D. 1987,
Phys. Rev. A, 35, 4354

\bibitem[Baym et al. (1971)] {Baym71}
Baym, G., Pethick, C., \& Sutherland, P. 1971,
Astrophys. J., 170, 299

\bibitem[Bell et al. (2013)] {Bell13}
Bell, N. F., Melatos, A., \& Petraki, K. 2013,
Phys. Rev. D, 87, 123507

\bibitem[Bertoni et al. (2013)] {Ber13}
Bertoni, B., Nelson, A. E., \& Reddy, S. 2013,
arXiv:1309.1721

\bibitem[Bertone \& Fairbairn (2008)] {Ber08}
Bertone, G., \& Fairbairn, M. 2008,
Phys. Rev. D, 77, 043515

\bibitem[Bramante et al. (2013)] {Kumar13}
Bramante, J., Fukushima, K., \& Kumar, J. 2013,
Phys. Rev. D, 87, 055012

\bibitem[Bramante et al. (2014)] {Kumar14}
Bramante, J., Fukushima, K., Kumar, J., \& Stopnitzky, E. 2014,
Phys. Rev. D, 89, 015010



\bibitem[Carriere et al. (2003)] {Car03}
Carriere, J., Horowitz, C. J., \& Piekarewicz, J. 2003,
Astrophys. J., 593, 463

\bibitem[Chabanat et al. (1997)] {Cha97}
Chabanat, E., Bonche, P., Haensel, P., Meyer, J., \& Schaeffer, R. 1997,
Nucl. Phys. A, 627, 710

\bibitem[Chang et al. (2010)] {Cha10}
Chang, S., Liu, J., Pierce, A., Weiner, N., \& Yavin, I. 2010,
JCAP, 08, 018

\bibitem[Chen (2011)] {Chen11a}
Chen, L. W. 2011,
Sci. China: Phys. Mech. Astro., 54, (Suppl. 1) s124

\bibitem[Chen (2011)] {Chen11}
Chen, L. W. 2011,
Phys. Rev. C, 83, 044308

\bibitem[Chen et al. (2001)] {Chen01}
Chen, L.-W et al. 2001,
Phys. Rev. C, 64, 064315

\bibitem[Chen \& Gu (2012)] {Chen12}
Chen, L. W., \& Gu, J. Z. 2012,
J. Phys. G, 39, 035104

\bibitem[Chen et al. (2010)] {Chen10}
Chen, L. W., Ko, C. M., Li, B. A., \& Xu, J. 2010,
Phys. Rev. C, 82, 024321

\bibitem[Cirigliano et al. (2013)] {Vin13}
Cirigliano, V.,  Graesser, M. L.,  Ovanesyan, G., \& Shoemaker, I. M. 2013,
arXiv:1311.5886

\bibitem[Cline \& Frey (2011)] {Cli11}
Cline, J. M., \& Frey, A. R. 2011,
Phys. Rev. D, 84, 075003

\bibitem[Colpi et al. (1986)] {Shapiro86}
Colpi, M., Shapiro, S. L., \& Wasserman, I. 1986,
Phys. Rev. Lett., 57, 2485


\bibitem[Lavallaz \& Fairbairn (2010)] {Lava10}
de Lavallaz, A., \& Fairbairn, M. 2010,
Phys. Rev. D, 81, 123521


\bibitem[Feng et al. (2011)] {Feng11}
Feng, J. L.,  Kumar, J.,  Marfatia, D.,  Sanford, D. 2011,
Phys. Lett. B, 703, 124-127

\bibitem[Feng et al. (2013)] {Feng13a}
Feng, J. L.,  Kumar, J., \&  Sanford, D. 2013,
Phys. Rev. D, 88,  015021

\bibitem[Feng \& Kumar et al. (2013)] {Feng13b}
Feng, J. L.,  Kumar, J.,  Marfatia, D., \&  Sanford, D. 2013,
arXiv:1307.1758.

\bibitem[Fitzpatrick \& Zurek (2010)] {Zur10}
Fitzpatrick, A. L., \& Zurek, K. M. 2010,
Phys. Rev. D, 82,  075004

\bibitem[Frandsen et al. (2011)] {Fra11}
Frandsen, M. T., Kahlhoefer, F., Sarkar, S., \& Schmidt-Hoberg, K. 2011,
JHEP, 1109, 128

\bibitem[Friedrich \& Reinhard (1986)] {Fri86}
Friedrich, J., \&  Reinhard, P.-G. 1986,
Phys. Rev. C, 33, 335



\bibitem[Gao et al. (2013)] {Gao13}
Gao, X., Kang, Z., \&  Li, T. 2013,
JCAP, 01, 021

\bibitem[Giuliani (2005)] {Giu05}
Giuliani, F. 2005,
Phys. Rev. Lett., 95, 101301

\bibitem[Goldman \& Nussinov (1989)] {Gold89}
Goldman, I., \& Nussinov, S. 1989,
Phys. Rev. D, 40, 3221

\bibitem[Gonzalez \& Reisenegger (2010)] {Gon10}
Gonzalez, D., \& Reisenegger, A. 2010,
A\&A, 522, A16

\bibitem[Gould (1987)] {Gould87}
Gould, A. 1987,
Astrophys. J., 321, 571

\bibitem[Gudmundsson et al. (1982)] {Gud82}
Gudmundsson, E. H., Pethick, C. J., \& Epstein, R. I. 1982,
Astrophys. J., 259, L19


\bibitem[Guver et al. (2014)] {Guv14}
Guver, T., Erkoca, A. E., Reno, M. H., \& Sarcevic, I. 2014,
JCAP, 05, 013



\bibitem[He et al. (2012)] {He12}
He, X.-G., Ren, B., \& Tandean, J. 2012,
Phys. Rev. D, 85, 093019

\bibitem[Ho \& Scherrer (2013)] {Ho10}
Ho, C. M., \& Scherrer, R. J. 2013,
Phys. Lett. B, 722,  341

\bibitem[Horowitz (2012)] {Horo12}
Horowitz, C. J. 2012,
arXiv:1205.3541



\bibitem[Iida \& Sato (1997)] {Iida97}
Iida, K., \& Sato, K. 1997,
Astrophys. J., 477, 294



\bibitem[Jamison (2013)] {Jami13}
Jamison, A. O. 2013,
Phys. Rev. D, 88, 035004





\bibitem[Kl\"{u}pfel et al. (2009)] {Klu09}
Kl\"{u}pfel, P., Reinhard, P.-G., B\"{u}rvenich, T.J., \& Maruhn, J. A. 2009,
Phys. Rev. C, 79, 034310

\bibitem[Kouvaris (2008)] {Chris08}
Kouvaris, C. 2008,
Phys. Rev. D, 77, 023006

\bibitem[Kouvaris (2012)] {Chris12}
Kouvaris, C. 2012,
Phys. Rev. Lett., 108, 191301

\bibitem[Kouvaris (2013)] {Chris13}
Kouvaris, C. 2013,
Adv. High Energy Phys., Vol. 2013, Article ID 856196

\bibitem[Kouvaris \& Tinyakov (2011)] {Chris11}
Kouvaris, C., \& Tinyakov, P. 2011,
Phys. Rev. Lett., 107, 091301

\bibitem[Kurylov \& Kamionkowski (2004)] {Kur04}
Kurylov, A., \&  Kamionkowski, M. 2004,
Phys. Rev. D, 69, 063503





\bibitem[Lattimer \& Prakash (2000)] {Latt00}
Lattimer, J. M., \& Prakash, M. 2000,
Phys. Rep.,  333, 121.

\bibitem[Lattimer \& Prakash (2001)] {Latt01}
Lattimer, J. M., \& Prakash, M. 2001,
Astrophys. J.,  550, 426

\bibitem[Lattimer \& Prakash (2004)] {Latt04}
Lattimer, J. M., \& Prakash, M. 2004,
Science, 304, 536

\bibitem[Li et al. (2008)] {LCK08}
Li, B. A., Chen, L. W., \&  Ko, C. M. 2008,
Phys. Rep., 464, 113

\bibitem[Link et al. (1999)] {Lat99}
Link, B., Epstein, R. I., \& Lattimer, J. M. 1999,
Phys. Rev. Lett., 83, 3362


\bibitem[Manchester et al. (2005)] {Man05}
Manchester, R. N., Hobbs, G. B., Teoh, A., \& Hobbs, M. 2005,
Astron. J. 129, 1993,
http://www.atnf.csiro.au/people/pulsar/psrcat/

\bibitem[McDermott et al. (2012)] {Yu12}
McDermott, S. D., Yu, H.-B., \& Zurek, K. M. 2012,
Phys. Rev. D, 85, 023519



\bibitem[Nagao \& Naka (2013)] {Nag13}
Nagao, K. I., \&  Naka, T. 2013,
Prog. Theor. Exp. Phys. B, 02, 043

\bibitem[Nobile et al. (2012)] {Del12}
Nobile, E. D., Kouvaris, C., Sannino, F., \& Virkajarvi, J. 2012,
Mod. Phys. Lett. A, 27, 1250108



\bibitem[Okada \& Seto (2013)] {Oka13}
Okada, N., \& Seto, O. 2013,
Phys. Rev. D, 88, 063506



\bibitem[Page et al. (2004)] {Pa04}
Page, D., Lattimer, J. M., Prakash, M., \& Steiner, A. W. 2004,
ApJS, 155, 623

\bibitem[Petraki \& Volkas (2013)] {Vol13}
Petraki, K. \&  Volkas, R. R. 2013,
Int. J. Mod. Phys. A, 28, 1330028

\bibitem[Press \& Spergel (1985)] {Press85}
Press, W.H., \& Spergel, D.N. 1985,
Astrophys. J., 296, 679



\bibitem[Savage et al. (2009)]{DAMA09}
Savage, C., et al. 2009,
JCAP, 0904, 010

\bibitem[Shapiro \& Teukolsky (1983)] {Sha83}
Shapiro, S. L., \& Teukolsky, S. A. 1983,
Wiley, New York, 1983, p. 569

\bibitem[Smith et al. (2007)] {Smith07}
Smith, M. C., et al. 2007,
Mon. Not. R. Astron. Soc., 379, 755


\bibitem[Wolszczan (1990)] {Wo90}
Wolszczan, A. 1990,
IAU Circ., 5073, 1

\bibitem[Wolszczan \& Frail (1992)] {Wo92}
Wolszczan, A., \& Frail, D. A. 1992,
Nature, 355, 145

\bibitem[Wolszczan (1994)] {Wo94}
Wolszczan, A. 1994,
Science, 264, 538


\bibitem[Xu et al. (2009)] {XuJ09}
Xu, J.,  Chen, L. W., Li, B. A., \&  Ma, H. R. 2009,
Astrophys. J, 697, 1549

\bibitem[Yakovlev \& Pethick (2004)] {Ya04}
Yakovlev, D. G., \& Pethick, C. J. 2004,
ARA\&A, 42, 169


\bibitem[Zhang \& Chen (2013)] {Zhang13}
Zhang, Z., \& Chen, L. W. 2013,
Phys. Lett. B, 726, 234

\bibitem[Zheng \& Chen (2012)] {Zheng12}
Zheng, H., \& Chen, L. W. 2012,
Phys. Rev. D, 85, 043013

\bibitem[Zheng et al. (2014)] {Zheng14}
Zheng, H., Zhang, Z., \&  Chen, L. W. 2014,
JCAP, 08, 011


\bibitem[Zurek (2014)] {Zur13}
Zurek, K. M. 2014,
Phys. Rep., 537, 91

\end{thebibliography}
\end{document}